\documentclass[twocolumn,superscriptaddress,pre,footinbib]{revtex4-1}

\newcommand{\tw}{\ensuremath{t_\mathrm{w}}\xspace}
\newcommand{\Tc}{\ensuremath{T_\mathrm{c}}\xspace}
\newcommand{\zc}{\ensuremath{z_\mathrm{c}}\xspace}

\newcommand{\ellJ}{\ensuremath{\ell_\text{J}}\xspace}
\usepackage[T1]{fontenc}
\usepackage[utf8]{inputenc}
\usepackage{graphicx}
\usepackage{amsmath}
\usepackage{amssymb}
\usepackage{color}
\usepackage{xspace}
\usepackage{hyperref}

\makeatletter
\newcommand*{\balancecolsandclearpage}{%
  \close@column@grid
  \cleardoublepage
  \twocolumngrid
}
\makeatother
\graphicspath{{./figures/}{./}}

\begin{document}
\title{On the Slowing Down of Spin Glass Correlation Length Growth:\\
  simulations meet experiments}

\author{Qiang Zhai}\affiliation{Texas Materials Institute, The University of Texas at Austin, Austin, Texas  78712, USA}
\author{V.~Martin-Mayor}\affiliation{Departamento de F\'\i{}sica
  Te\'orica, Universidad Complutense, 28040 Madrid, Spain}
\affiliation{Instituto de Biocomputaci\'on y F\'{\i}sica de Sistemas Complejos
  (BIFI), 50018 Zaragoza, Spain}
\author{Deborah L. Schlagel}
\affiliation{Division of Materials Science and Engineering, Ames Laboratory,
  Ames, Iowa, 50011, USA}
\author{Gregory G. Kenning}
\affiliation{Department of Physics, Indiana University of Pennsylvania,
  Indiana, Pennsylvania, 15705, USA}
\author{Raymond~L.~Orbach}
\affiliation{Texas Materials Institute, The University of Texas at Austin,
  Austin, Texas  78712, USA}
%%%%%%%%%%%%%%%%%%%%%%%%%%%%%%%%%%%%%%%%%%%%%%%%%%%%%%%%%%%%%%%%%%%%%%

\begin{abstract}
  The growth of the spin-glass correlation length has been measured as a
  function of the waiting time $\tw$ on a single crystal of CuMn (6 at.\%),
  reaching values $\xi\sim 150$ nm, larger than any other glassy
  correlation-length measured to date. We find an aging rate $\mathrm{d}\ln\,
  \tw/\mathrm{d}\ln\,\xi$ larger than found in previous measurements, which
  evinces a dynamic slowing-down as $\xi$ grows. Our measured aging rate is
  compared with simulation results by the Janus collaboration. After
  critical effects are taken into account, we find excellent agreement with the
  Janus data.
\end{abstract}

\date{\today}
\maketitle

%%%%%%%%%%%%%%%%%%%%%%%%%%%%%%%%%%%%%%%%%%%%%%%%%%%%%%%%%%%%%%%%%%%%%%

%%%%%%%%%%%%%%%%%%%%%%%%%%%%%%%%%%%%%%%%%%%%%%%%%%%%%%%%%%%%%%%%%%%%%%%%%%%
\section{Introduction.}\label{sect:Introduction}
The accuracy provided by SQUIDs in measurements of the response to an
externally applied magnetic field put spin-glasses in a privileged status
among glassy systems~\cite{cavagna:09} in at least two respects.  First, we
know that their sluggish dynamics originates in a \emph{bona fide} phase
transition at a critical temperature $T_\mathrm{c}$, separating the
paramagnetic phase from the low-temperature glassy
phase~\cite{gunnarsson:91}. Second, the suspected mechanism for the dynamic
slow-down, namely the increasing size of the cooperative
regions~\cite{adam:65}, has been confirmed experimentally~\cite{joh:99}. The
size of these cooperative regions, the so called spin-glass correlation length
$\xi$, was found to be as large as $\xi\approx 80$ nm (much larger than found
to date in other glassy systems, glycerol for instance~\cite{albert:16}).

In the typical set-up, the spin glass is rapidly cooled from high temperatures
to a working temperature $T<T_\mathrm{c}$, where it relaxes for a waiting time
$\tw$.  In principle, the growth of the correlation length $\xi(\tw)$ is
unbounded in the spin-glass phase (however, finite crystallite sizes play
a role, see below). Much attention has been paid to the (renormalized)
aging-rate
\begin{equation}\label{eq:aging-rate}
\zc(T,\xi)= \frac{T}{T_\mathrm{c}} \frac{\mathrm{d}\,\ln\,
  \tw}{\mathrm{d}\, \ln \,\xi}\,.
\end{equation}
The renormalizing factor $T/T_\mathrm{c}$ makes $\zc(T,\xi)\approx
\zc(\xi)$~\footnote{General theoretical arguments suggest that $\zc$
  is also $\xi$-independent at exactly $T=\Tc$~\cite{zinn-justin:05}.
  Only if $\zc$ is $\xi$-independent we have a power-law scaling $\tw
  \propto \xi^{\Tc\zc/T}$. If $\zc$ grows with $\xi$, as we find here,
  we encounter a dynamics slower than a power-law (for instance, an
  activated dynamics with free-energy barriers $\Delta\propto\xi^\Psi$
  for some $\Psi>0$).}. Hence, Eq.~\eqref{eq:aging-rate} can be
rephrased as $t_w^{\text{eff}}\approx \tau_0
\exp[(\Delta(\xi)-E_z(H))/k_BT]$ where
$\tau_0=\hbar/(k_\mathrm{B}\Tc)$ is the exchange time, $E_z$ is the
Zeeman energy and $\Delta(\xi)$ is a free-energy barrier.

In fact, values of $\zc$ have been found to vary from system to system.  For a
bulk, polycrystalline sample, of CuMn 6 at.\%; Joh et al. \cite{joh:99} found
at a reduced temperature $T/\Tc=0.89$, $\zc=5.917$.  For a polycrystalline
bulk thiospinel, Joh et al. found at a reduced temperature of $T/\Tc=0.72$,
$\zc=7.576$.  There is no way of knowing the crystallite size in these ``bulk"
measurements, but they were certainly larger than the thin film thicknesses of
Zhai et al. \cite{zhai:17}.  Zhai et al. found, for CuMn 11.7 at.\% thin films
at reduced temperatures of $T/\Tc=0.43,~ 0.59,~ 0.78$, $\zc=9.62$. 
Working at $T/\Tc=0.95$, Kenning et al. \cite{kenning:18} obtained $\zc=6.80$
in a bulk polycrystalline CuMn 5 at.\% sample.

Some hints to classify these apparently conflicting results can be
found in a recent large-scale numerical simulation by the Janus
collaboration~\cite{janus:18} (using the custom built computer Janus
II~\cite{janus:14}). They computed $\xi$ in a time range $10^{-12} $ s
$\leq\tw\leq 0.1$ s for temperatures $0.5\leq T/\Tc \leq 1$. In fact,
$\xi$ varied by a larger factor in the simulation than in experiments:
close to $\Tc$, from $\xi\sim a_0$ to $\xi\sim 17\, a_0$ ($a_0$ is the
typical distance between magnetic moments). Yet, the maximum $\xi/a_0$
reached in the simulations was smaller than experiment by a factor of
approximately 10.

The Janus simulation evinced different behaviors at $\Tc$ and
at $T<\Tc$~\cite{janus:18}, according to the value
of the crossover
variable:
\begin{equation}\label{eq:x-def}
x(\tw,T)=\ellJ(T)/\xi(\tw,T)\, , 
\end{equation}
 where $\ellJ(T)$ is the Josephson length~\cite{josephson:66}.
For $x\ll 1$ we have $T<\Tc$ behavior, while for $x\gg 1$ we find critical
scaling.  Because $\ellJ(T)$ diverges at $\Tc$ as $\ellJ(T)\propto
1/(\Tc-T)^\nu$ , $\nu=2.56(4)$~\cite{janus:13}, the $\xi(\tw)$ needed to
demonstrate low-temperature behavior, i.e. $x\ll 1$, grows enormously upon
approaching $\Tc$.  For $x\ll 1$, $\zc$ grows with $\xi$, but it is
$T$-independent~\cite{janus:18}. Furthermore, a mild extrapolation from
$\zc(\xi=12\, a_0)$ to $\zc(\xi=38\, a_0)$ ~\cite{janus:18} is compatible with
the thin-film value $\zc=9.62$~\cite{zhai:17} (the film width was $\sim
38\,a_0$). For $x \gg 1$, the $\xi$-independent $\zc(T=\Tc)=6.69\pm
0.06$~\cite{janus:18} agrees with the CuMn result at $T=0.95\Tc$, $\zc=6.80$
\cite{kenning:18}.

However, in spite of the just quoted agreement between experimental
results and the Janus simulations, the reader might worry because CuMn
is a Heisenberg spin-glass, while the Janus collaboration simulates
the Ising-Edwards-Anderson model. In fact, there is theoretical ground
for the success of the Ising spin-glass simulations: small
anisotropies such as Dzyaloshinsky-Moriya interactions~\cite{bray:82}
are present in any spin-glass sample. These interactions, though tiny,
extend over dozens of lattice spacings, which magnifies their
effect. In fact, we know that Ising is the ruling universality class
in the presence of coupling anisotropies~\cite{baityjesi:14} (the
effect of anisotropies, even if negligible at small $\xi$, is strongly
enhanced when $\xi$ grows~\cite{amit:05}), which probably explains why
high-quality measurements on GeMn are excellently fit with Ising
scaling laws~\cite{guchhait:17}.

Here, we report measurements of $\xi(\tw)$ on a single crystal of CuMn (6
at.\%), at $T=0.886\, \Tc$ and for times $2\times 10^3$ s $\leq \tw\leq
8\times 10^4$ s. In the absence of crystallites limiting $\xi$ to the
crystallite size ($\sim 80$ nm, typically), we reach $\xi \sim 150$ nm, a
world record in a glassy phase (and, certainly, in the low-temperature regime
$x\ll 1$). Our measured aging rate $z_c = 12.37\pm 1.07$ is the largest ever
measured in a spin-glass, in a dramatic demonstration of the dynamic
slowing-down with growth of $\xi$~\cite{janus:18}. We are also able to reproduce
our experimental results by means of a simple extrapolation of the Janus
simulations~\cite{janus:18}.

The layout of the remaining part of this paper is as follows. In
Sect.~\ref{sect:sample} we provide details about our single-crystal
sample.  Our experimental protocol is explained in
Sect.~\ref{sect:protocol}. Our extrapolation from the Janus
simulations is confronted with the experimental results in
Sect.~\ref{sect:extrapolations}. We present our conclusions in
Sect.~\ref{sect:conclusions}. The manuscript ends with a number of
appendices were more technical details are given.

\section{Sample preparation.}\label{sect:sample}  
The Cu$_{94}$Mn$_{6}$ sample was prepared using the Bridgman
method.  The Cu and Mn were arc melted several times in an Argon
environment and cast in a copper mold.  The ingot was then processed
in a Bridgman furnace.  Both XRF (X-ray fluorescence) and optical
observation showed that the beginning of the growth is a single phase.
More details can be found in Appendix~\ref{app:sample}.

\section{Experimental protocol.}\label{sect:protocol} We follow the method introduced by Joh et al.~\cite{joh:99} for the
extraction of $\xi(\tw)$, standard in experimental work (see
e.g.~\cite{Wandersman_2008,PhysRevB.75.214415}) and studied
theoretically~\cite{janus:17b}.

Specifically, the CuMn sample was quenched from 70 K to 28 K in zero magnetic
field ($T_g = 31.5$ K as determined from the temperature at which the remanence disappeared). This measurement temperature was determined by two factors.  To have measured at a higher temperature would have increased the Josephson length, increasing $x(t_w,T)$ according to Eq. (2).  It was important to keep $x(t_w,T)$ as small as possible in order to have $T<T_c$ behavior.  In addition, the signal to noise diminishes as the measuring temperature $T$ increases.  The lower $T$, the slower the dynamics. The working temperature $T = 28$ K was chosen so as to keep the measurements within laboratory time scales.

\begin{figure*}
\centering
  \includegraphics[width=0.32\textwidth]{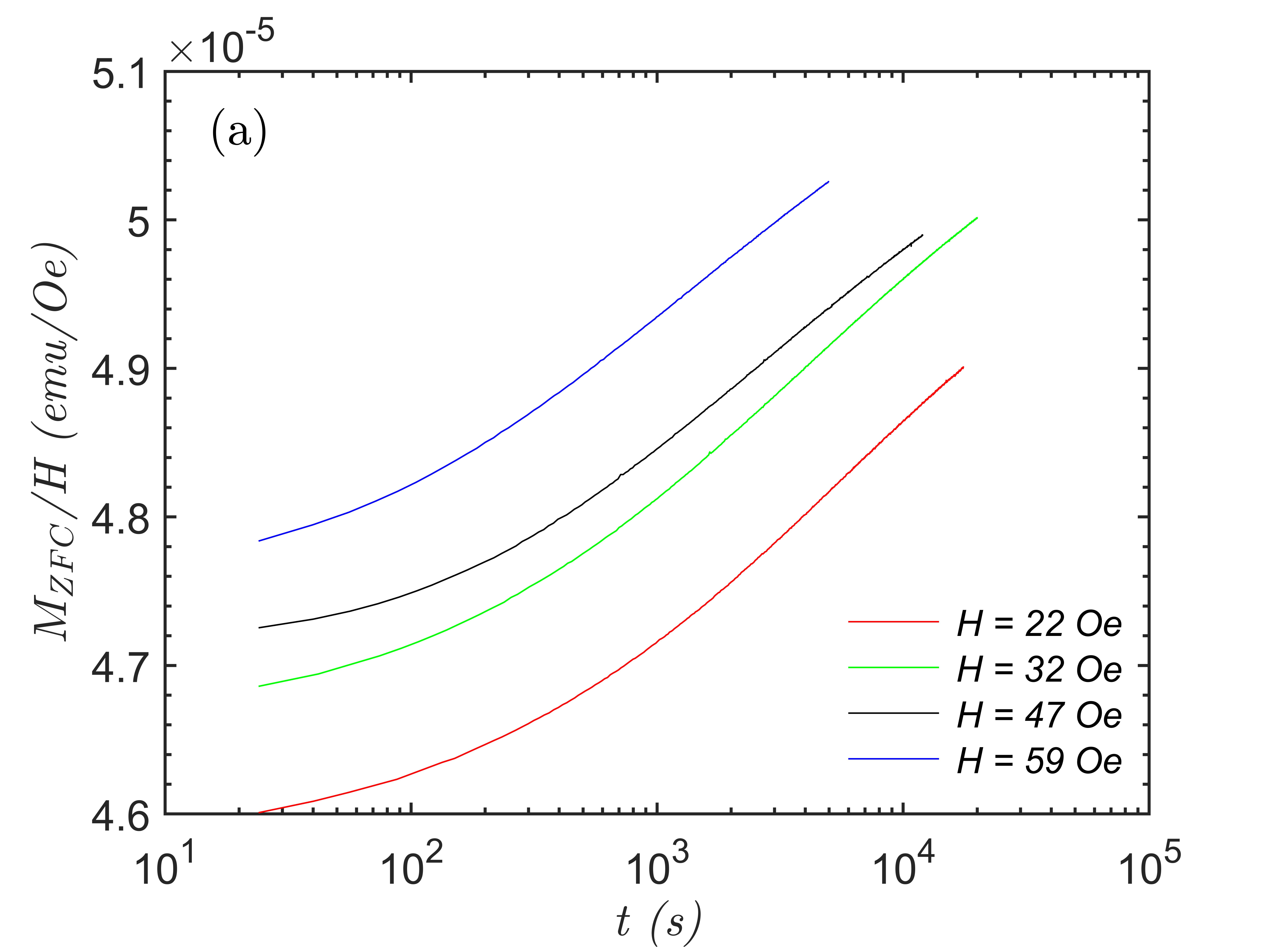}
  \includegraphics[width=0.32\textwidth]{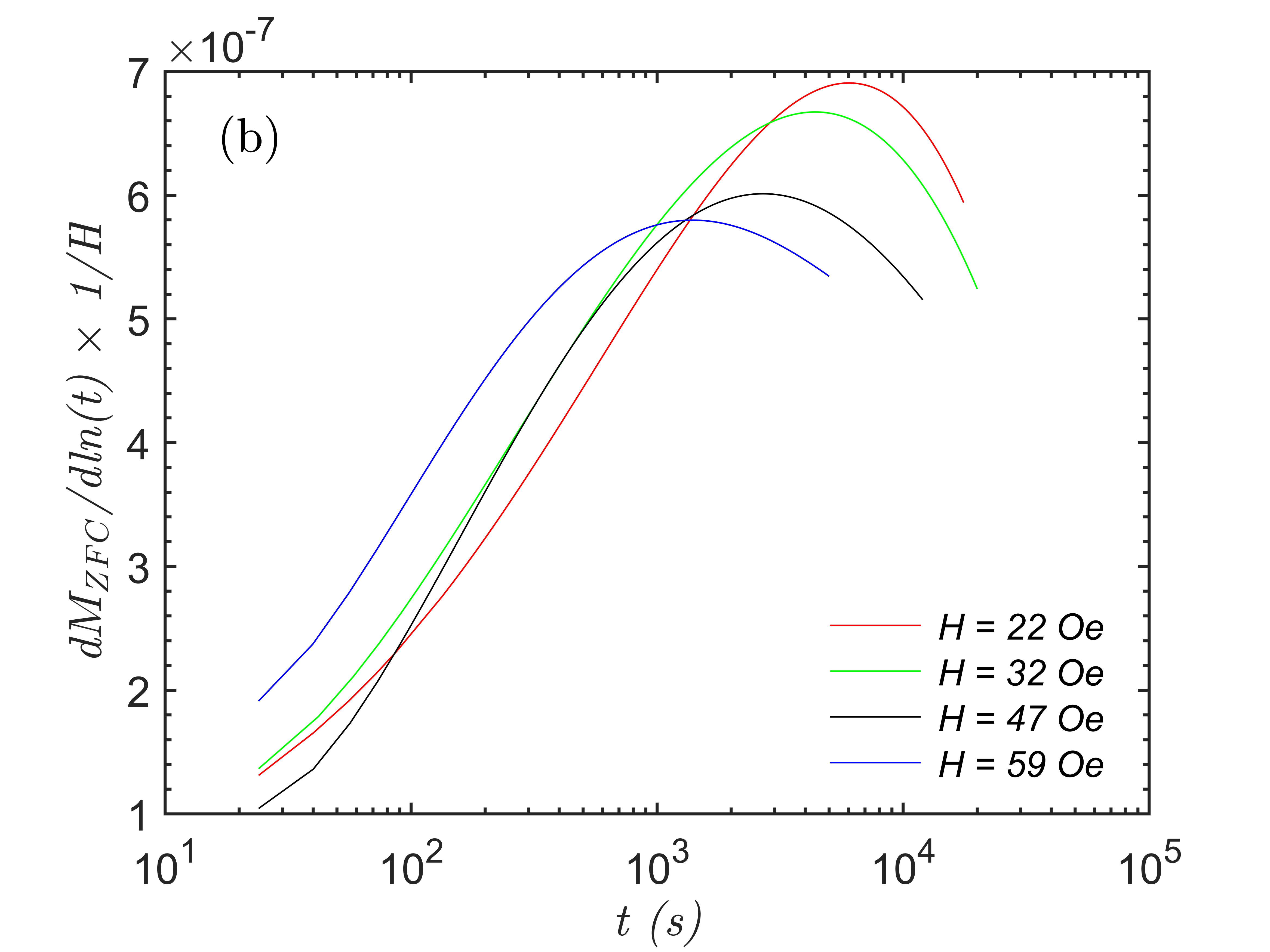}
  \includegraphics[width=0.32\textwidth]{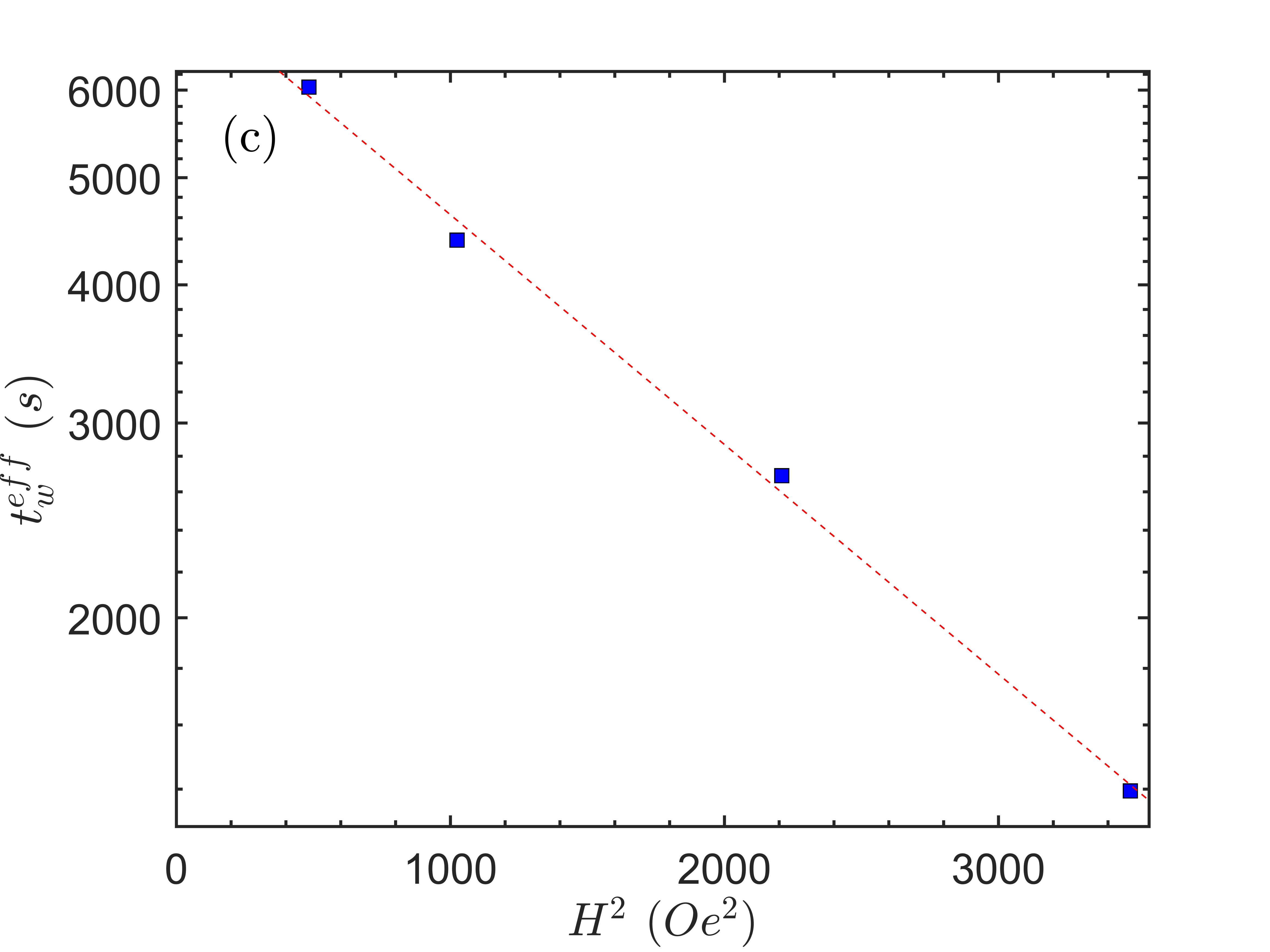}
\caption{\label{fig:intermediate} A representative set of data.  The three figures are for a waiting time $t_w$ = 10 ks. (a) A plot of the measured zero field susceptibility, $M_{ZFC}/H$, as a function of time.  (b) The response function, $S(t) = d\,(M_{ZFC}/H)/d\,(\ell n\,t)$ as a function of time for varying values of the applied magnetic field $H$, the peak of which defines $t_w^{\text {eff}}$. (c) A plot of $\ell n\,t_w^{\text {eff}}$ vs $H^2$.}
\end{figure*}

The system was aged for a time
$t_w$ after the temperature has been stabilized, then a magnetic field $H$ was
applied, and 24 s after the field stabilized, the zero-field magnetization,
$M_{\text {ZFC}}(t,T)$, was recorded ($t$ is the time elapsed since the
magnetic field was switched on).  In this set of experiments, $t_w$ was set as
2 000, 2 750, 3 420, 5 848, 10 000, 20 000, 40 000, and 80 000 seconds, with
magnetic fields of 20, 32, 47, and 59 Oe.  The latter are used for the
magnetic field dependence of the effective waiting time, $t_w^{\text {eff}}$
as determined from the time for the relaxation function to reach its maximum
as a function of $\ln\, t$,
\begin{equation}
  S(t)={\frac {{\mathrm d}\,M_{\text {ZFC}(t)}}{\mathrm{d}\,\ln\,t}}\,.
\end{equation}
Note that the effective waiting time $t_w^{\text {eff}}$ where $S(t)$
attains its maximum depends on the
applied magnetic field, because the Zeeman effect lowers the free energy barrier
heights.  This results in a shift of the peak in $S(t)$ (its maximum
$t_w^{\text {eff}}$):
\begin{equation}
\Delta_{\text {max}}-N_c\,\chi\,H^2=k_BT\,\ln\, t_w^{\text {eff}}-k_BT\,\ln\, \tau_0~~,
\end{equation}
where $N_c$ is the number of spin glass correlated spins, $\chi$ is
the spin glass field-cooled susceptibility per spin [$M_{\text
    {FC}}/(NH)$, with $N$ the total number of Mn spins in the sample],
and $\tau_0$ is an effective exchange time
$\tau_0\sim\hbar/(k_\text{B}T_g)$.  The beauty of this expression is
that $N_c$ can be determined from Eq. (3) from measurement of the peak
position of $S(t)$ as a function of $H^2$, and from other known values
of the parameters. A representative set of data is exhibited in
Fig.~\ref{fig:intermediate}. Our $t_w^{\text {eff}}$ are in
Table~\ref{Tab:tweff}.

\begin{table}
	\caption{Effective waiting time $t_w^{\text{eff}}$ extracted in ZFC (zero field cooled ) magnetization aging experiments.\label{Tab:tweff}}
	\setlength{\tabcolsep}{5pt}
	\begin{tabular}{c c c c c}
		\hline
		\hline
		 $t_w (s)$   & H = 22 Oe & H = 32 Oe & H = 47 Oe & H = 59 Oe\\
		\hline
		 2 000 &1 463&1 161&727\footnote{measured in 50 Oe}&593\\
	     2 750 &1 924&1 599&1 009&696\\
	     3 420 &2 395&1 832&1 069&726\\
	     5 848 &3 860&2 865&1 615&1 058\\
	     10 000 &6 038&4 390&2 689&1 395\\
	     20 000&11 978&8 073&4 047&2 104\\
	     40 000&21 710&14 601&6 838&3 451\\
	     80 000&41 748&26 215&11 467&5 266\\
	     \hline
	     \hline
	\end{tabular}
\end{table}

Knowing $N_c$, the correlation length $\xi$ can be generated from the
relationship~\footnote{The reader will note that the right-hand side of Eq. (5) could be modified by a prefactor of order 1.This is why we are using an approximate sign in the equation, rather than an equal sign. However, the comparison with the simulations turns out to be satisfactory by assuming that the prefactor is exactly one. It is well possible that carrying out our program from future experiments of increased accuracy will require a more precise determination of this prefactor.},
\begin{equation}\label{eq:Nc-xi}
N_c\approx \bigg({\frac {\xi}{a_0}}\bigg)^{d_{\text{f}}}\,,
\end{equation}
where $d_{\text{f}}$ is the fractal dimension equal to $d_{\text{f}}=d-\theta/2$ ($d=3$ is the
space dimension, while $\theta$ is the so-called replicon
exponent~\cite{janus:17b}). Because at the correlation lengths of interest
$\theta\approx 0.3$, the approximation $d_{\text{f}}\approx d$ made in previous work
(Ref.~\cite{joh:99}, for instance) does not introduce a significant error.

\begin{figure}[t]
\centering
\includegraphics[width=\columnwidth]{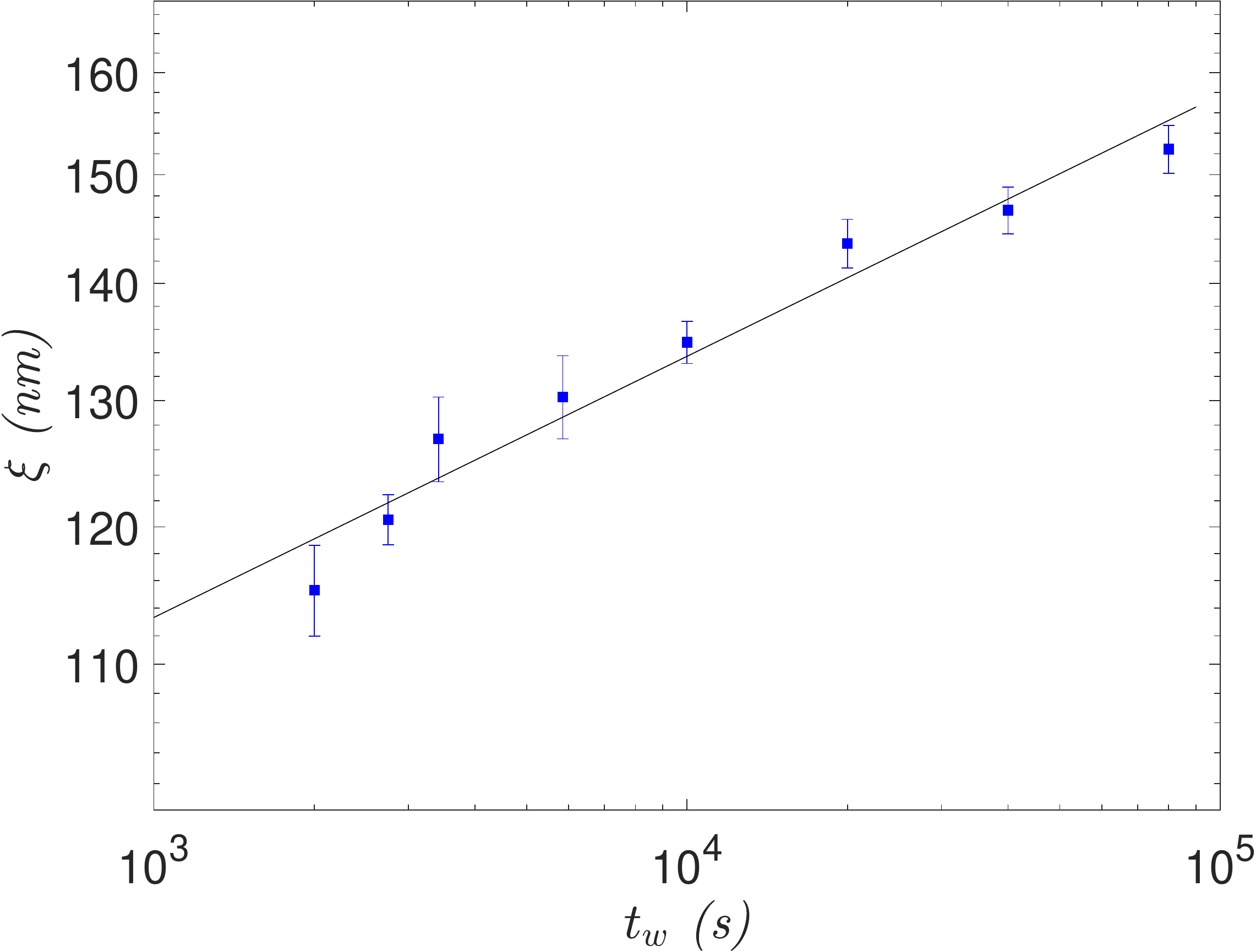}
\caption{\label{fig:xi-log-linear-fit} $\xi(t_w,T)$ as a function of waiting time,
  $t_w$ at a measuring temperature $T=28$ K (the transition temperature is
  $\Tc\approx 31.5$ K).  The straight line is a fit to $\ln \,\tw=
  (\zc\Tc/T)\ln\,\xi + constant$, recall Eq.~\eqref{eq:aging-rate}, yielding
  $z_c=12.37 \pm 1.07$.}
\end{figure}
In fact, the exponent $\theta$ has a small dependency on
$\xi$~\footnote{Amusingly, although the droplets
  model~\cite{mcmillan:83,bray:87,fisher:86} and the Replica Symmetry Breaking
  (RSB) theory~\cite{marinari:00} differ in their expectation for
  $\theta(\xi\to\infty)$ (the droplets prediction is $\theta(\xi\to\infty)=0$
  and $d=d_{\text{f}}$, while RSB expects $\theta(\xi\to\infty)>0$ and $d_{\text{f}}<d$), the
  two theories quantitatively agree in their predicted behavior for
  $\theta(\xi)$ in our range of $\xi$~\cite{janus:18}.}.  We have
solved this problem by taking the exponent $\theta(\xi)$ from
Ref.~\cite{janus:18} and then solved for $\xi$ in Eq.~\ref{eq:Nc-xi}
self-consistently (see Appendix~\ref{app:replicon}). The appropriate value of $\theta$ turns
out to be $\theta\approx 0.34$. The outcome of this analysis is shown in
Fig.~\ref{fig:xi-log-linear-fit}. The estimated Josephseon length at our working
temperature is $\ellJ=21.82\, a_0$ ($a_0=0.64$
nm in our sample), see Ref.~\cite{janus:18} and Appendix~\ref{app:Josephson}. Hence, the crossover variable in our experiment is in the
range $0.091 \leq x\leq 0.12$, so that we can be reasonably sure to be free from
critical effects. The resulting aging-rate is $z_c=12.37 \pm 1.07$.
Comparing with previous values of $z_c$, obtained in experiments reaching a
smaller $\xi(t_w,T)$~\cite{joh:99,zhai:17,kenning:18}, this is the largest
aging rate ever measured in a spin glasses, which shows that the growth of $\xi$
is indeed slowing down with increasing $\xi$.

\begin{figure}[t]
\centering
\includegraphics[width=\columnwidth]{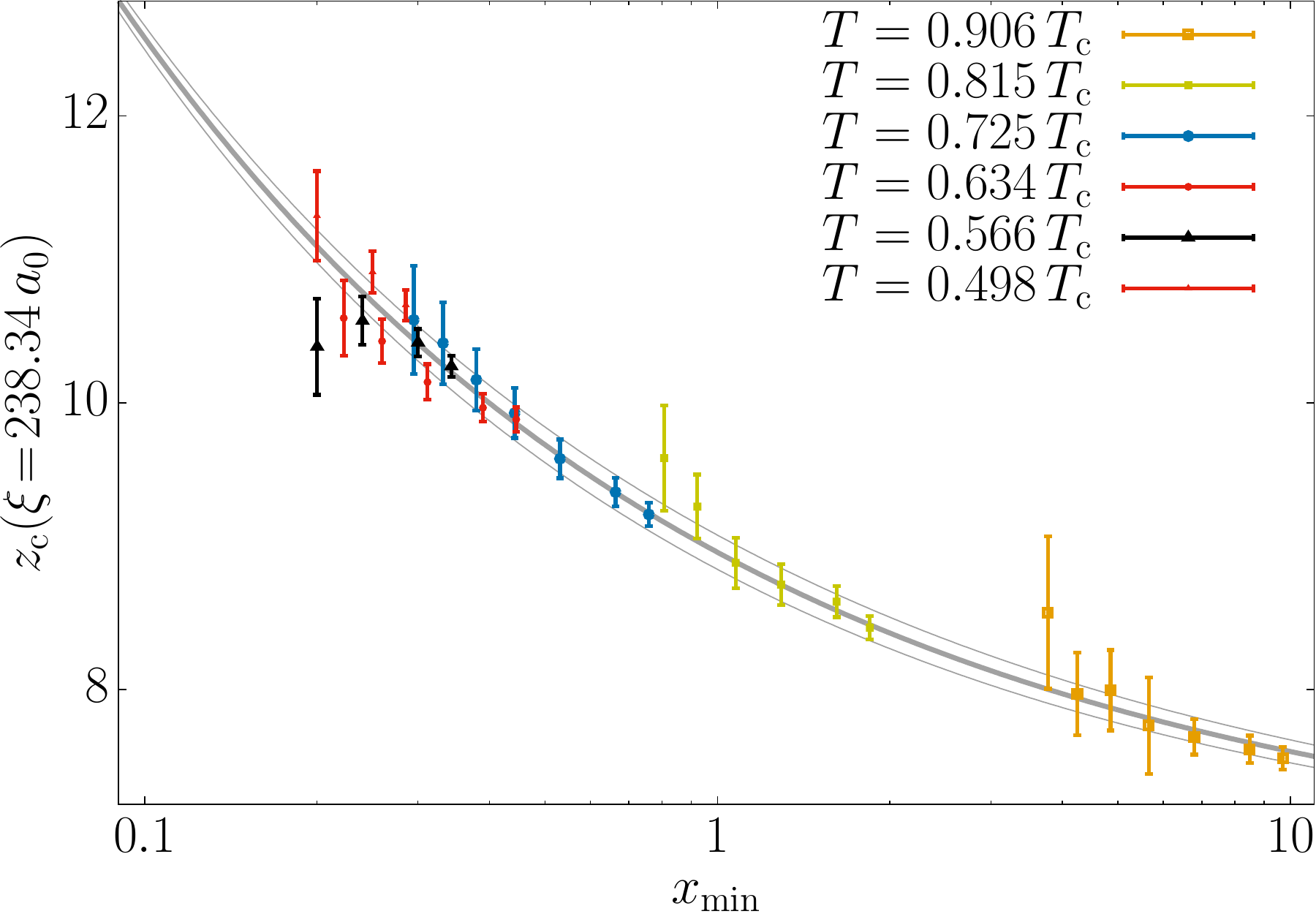}
\caption{\label{fig:zc-xitarget238.4} The estimates from different
  temperatures and minimal correlation lengths for the aging rate at
  $\xi_{\text{target}}=238.34~a_0$ (our largest) are a simple function
  of the crossover variable $x_\text{min}=\ellJ(T)/\xi_{\text{min}}$,
  see Eq.~\eqref{eq:well-organized-data}. The central black line is a
  fit to Eq.~\eqref{eq:well-organized-data} with figure of merit
  $\chi^2/{\text {dof}}=24.5/30$ [dof = degrees of freedom. The fit
    generates the exponent
    $\beta(\xi_{\text{target}}=238.34~a_0)=0.41$, the dependency on
    $\xi_{\text{target}}$ of exponent $\beta$ turns out to be
    small]. The upper (lower) black line is a fit to the data plus
  (minus) the error bar. The estimates of $\zc$ for the different
  $(T,\xi_{\text{min}})$ were obtained by applying
  Eq.~\eqref{eq:convergent-ansatz} to the data in Table III of the SM
  for Ref.~\cite{janus:18} (see Appendix~\ref{app:extrapolation} for
  details).}
\end{figure}

\section{Extrapolations from simulations.}\label{sect:extrapolations}

The main problem to overcome is the crossover between critical scaling
and the $T<\Tc$ Physics. Indeed, the largest correlation length
reached in the simulations is $\xi=17.3\, a_0$ at
$T=0.905\Tc$~\cite{janus:18}, which results in a very large cross-over
variable $x=1.96$. Much smaller values of $x$ were reached in the
simulations, but at lower $T$~\cite{janus:18}. Therefore, we need to
consider the full data-set for $T<\Tc$ in Table III of the SM for
Ref.~\cite{janus:18}. We shall only outline our analysis here and
refer the reader to Appendix~\ref{app:extrapolation} for full details. To
ease comparison with~\cite{janus:18} , we give $\xi$ in units of $a_0$
from now on (recall that $a_0=0.64$ nm for our sample).

We should mention that two possibilities were considered in
Ref.~\cite{janus:18} for extrapolating the simulation's $\zc$ to larger values
of $\xi$. One was Saclay's ansatz for the crossover to activated
dynamics~\cite{bouchaud:01,berthier:02} which, however, yields too-high a
$\zc$~\cite{janus:18} when applied to the thin-film
experiments~\cite{zhai:17}. Therefore, we focus on the
convergent ansatz for extrapolating $\zc$ to
correlation length $\xi_{\text{target}}$ by taking into account only
data with $\xi\geq \xi_{\text{min}}$ ($\hat\omega=0.35$)~\cite{janus:18}
\begin{equation}\label{eq:convergent-ansatz}
\zc(T,\xi_{\text{target}},\xi_{\text{min}})=\frac{T}{\Tc}\Big[z_\infty(T,\xi_{\text{min}})+\frac{A(T,\xi_{\text{min}})}{\xi_{\text{target}}^{\hat\omega}}\Big]\,.
\end{equation}
Now, when applying Eq.\eqref{eq:convergent-ansatz} to any
$\xi_{\text{target}}$, we end-up with as many predicted aging-rates as pairs
of $(T,\xi_{\text{min}})$ were considered in the simulations. Fortunately,
these many predictions, see Fig.~\ref{fig:zc-xitarget238.4}, can be nicely
organized as a function of the crossover variable
$x_{\text{min}}=\ellJ(T)/\xi_{\text{min}}$~\footnote{Note that
  Eq.~\eqref{eq:well-organized-data} correctly predicts $\zc(\Tc)=6.69$,
  because $x=\infty$ at $\Tc$, for all $\xi$.}:
\begin{equation}\label{eq:well-organized-data}
\zc(T,\xi_{\text{target}},\xi_{\text{min}})=6.69+\frac{\alpha(\xi_{\text{target}})}{x_{\text{min}}^{\beta(\xi_{\text{target}})}}\,\,.
\end{equation}
Thus, our final
extrapolation at $T=28$ K is
\begin{equation}\label{eq:zc-final}
  \zc(\xi_\text{target})=6.69+
  \frac{\alpha(\xi_{\text{target}})}{x_{\text{target}}^{\beta(\xi_{\text{target}})}}\,,\ 
  x_{\text{target}}=\frac{\ellJ(28\text{K})}{\xi_{\text{target}}}\,,
\end{equation}
[$\alpha(\xi_{\text{target}})$ and
$\beta(\xi_{\text{target}})$ come from the fit 
to Eq.~\eqref{eq:well-organized-data}, recall
Fig.~\ref{fig:zc-xitarget238.4}]. We obtain in this way
\begin{equation}\label{eq:extrapolated-zc}
\zc(180.26\,a_0)=11.94\pm0.08\,,\ \zc(238.34\,a_0)=12.76\pm 0.08\,.
\end{equation}
Both extrapolations are in excellent agreement with the
experimental result $z_c=12.37 \pm 1.07$ from
Fig.~\ref{fig:xi-log-linear-fit} (roughly speaking, $z_c=12.37 \pm 1.07$
is an average of $\zc(\xi)$ in the range $180.26\, a_0\leq \xi\leq 238.34\, a_0
$).

\begin{figure}[t]
\centering
\includegraphics[width=\columnwidth]{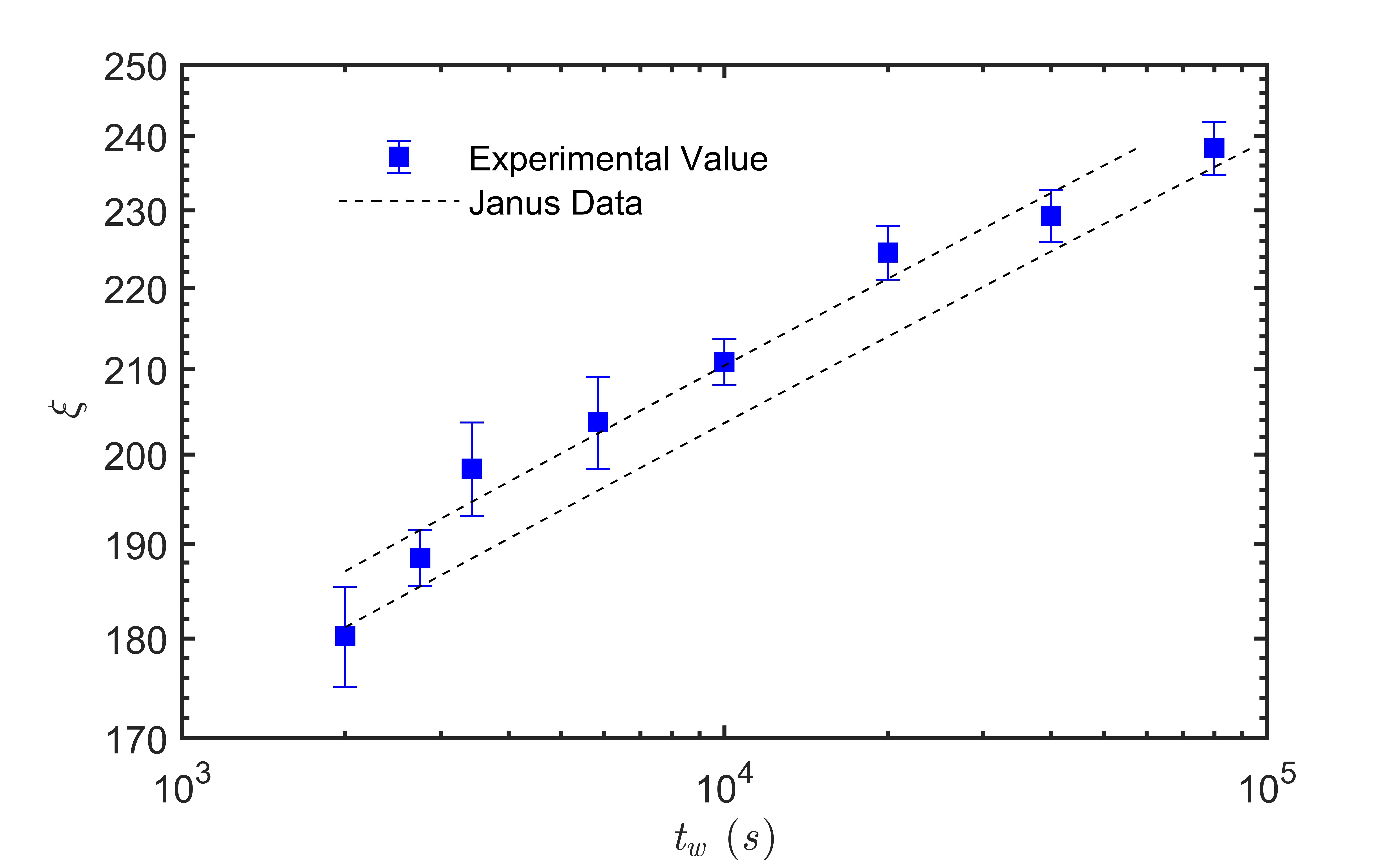}
\caption{\label{fig:xi-tw-experiment-extrapolations} The experimental
  correlation length from Fig.~\ref{fig:xi-log-linear-fit}, as measured in units of the
  average distance between magnetic moments $a_0=0.64$ nm,  is shown as a
  function of the waiting time. The two continuous lines are obtained from our
  extrapolations from the simulations by the Janus
  collaboration~\cite{janus:18}, recall
  Eqs.~(\ref{eq:zc-final},\ref{eq:direct-comparison}).
 The two lines are the two extremal curves compatible with the initial
 condition taken from our experiment, $\xi^*(\tw^*=2750\,\text{s})=(188.5\pm 3)\,a_0$.}
\end{figure}
We stress that the extrapolations~\eqref{eq:extrapolated-zc} took no input
from the experiment other than the values of $\xi_{\text{target}}$. However,
by recalling
[see Eq.~\eqref{eq:aging-rate}]
\begin{equation}\label{eq:direct-comparison}
\ln\,\tw -\ln\,\tw^*=\int_{\ln\xi^*}^{\ln\xi}\,\mathrm{d}(\ln\xi^\prime)\, \frac{\Tc}{T}\zc(\xi^\prime)\,,  
\end{equation}
and borrowing the initial condition 
$\xi^*(\tw^*=2750\,\text{s})$ from the experiment, we obtain
a fairly satisfactory comparison between our experiment and our extrapolations
from the Janus simulations in Fig.~\ref{fig:xi-tw-experiment-extrapolations}. We note as well that the
initial condition $\xi^*(\tw=2000\,\text{s})$ from the experiment, afflicted by larger errors
and short-time systematic effects, produces similar extrapolated curves.

\section{Conclusions.}\label{sect:conclusions}
We have reported an experimental measurement of the
spin-glass correlation length in a single-crystal sample of CuMn (6
at.\%).  Our experiment is free from two systematic effects
encountered in previous work: (i) the growth of the correlation length
is not hampered by the sample geometry (neither
crystallites~\cite{joh:99} or the film-thickness~\cite{zhai:17}) and
(ii) our results are representative of the low-temperature phase
(i.e. they are not contaminated by critical scaling), as shown by the
small value of the cross-over variable we reach [recall
  Eq.~\eqref{eq:x-def}]. We report the largest spin-glass correlation
length ever measured in a glassy phase. Our aging rate is also the
largest to date (at least as measured in a spin-glass). We thus
confirm the slowing down as $\xi$ grows that was suggested by the
simulations of the Janus collaboration~\cite{janus:18}. Furthermore,
we have been able to reproduce our experimental results by means of a
simple extrapolation of the Janus results. We believe this relation
between simulations and experiment opens new opportunities in
condensed matter physics. The complementary contributions allow
exploration of phenomena, especially in complex systems, with the
particular insights of each partner fueling the interpretation and
development of the other. This paper is the beginning of this new
research relationship.
%%%%%%%%%%%%%%%%%%%%%%%%%%%%%%%%%%%%%%%%%%%%%%%%%%%%%%%%%%%%%%%%%%%%%%%%%%%
\begin{acknowledgments}
We warmly thank the Janus collaboration for allowing us to reanalyze
their results. We also thank L.A. Fern\'andez for his help with figure
preparation. We thank helpful discussions with S. Swinnea about sample
characterization. This work was partially supported by the US
Department of Energy, Office of Basic Energy Sciences, Division of
Materials Science and Engineering, under award No. DE-SC0013599 and
contract No. DE-AC02-07CH11358, and by the Ministerio de Econom\'ia,
Industria y Competitividad (MINECO) (Spain) through Grants
No. FIS2015-65078-C2 and PGC2018-094684-B-C21 (contracts partially
funded by FEDER).
\end{acknowledgments}
  
\appendix
\section{Sample Preparation}\label{app:sample}

Crystal growth and sample preparation was carried out by the Materials
Preparation Center (MPC) of the Ames Laboratory, USDOE.  Cu from Luvata
Special Products (99.99 wt \% with respect to specified elements) and
distilled Mn from the MPC (99.93 wt\% with respect to all elements) was arc
melted several times under Ar and then drop cast in a water chilled copper
mold.  The resulting ingot was placed in a Bridgman style alumina crucible and
heated under vacuum in a resistance Bridgman furnace to 1050$^0$C, just above
the melting point. The chamber was then backfilled to a pressure of 60 psi
with high purity argon to minimize the vaporization of the Mn during the
growth.  The ingot was then further heated to 1300$^0$C and held for one hour to
ensure complete melting and time for the heat zone to reach a stable state.
The ingot was withdrawn from the heat zone at a rate of 3mm/hr.  About 1/3 of
the crucible stuck to the alloy.  The ingot was finally freed after
alternating between hitting with a small punch and hammer and submerging in
liquid nitrogen.

\begin{figure}[t] 
	\centering
	\resizebox{3in}{2in}{\includegraphics{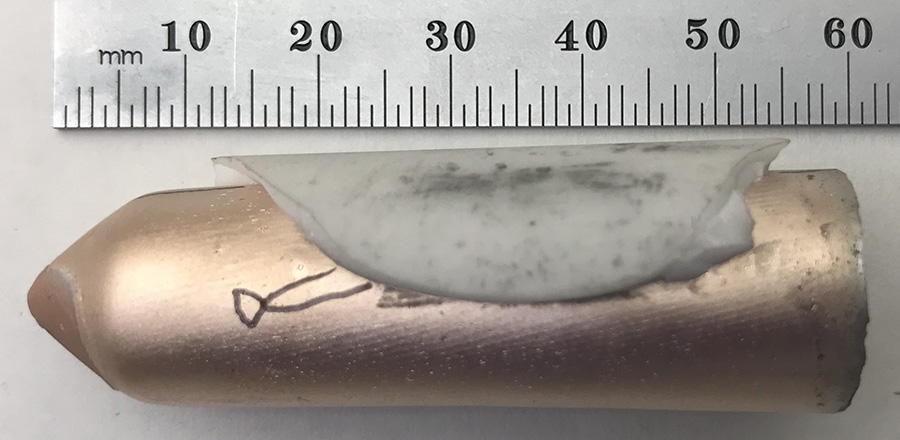}}
	\caption[]{\label{fig:ingot}The as-grown crystal with part of the alumina crucible still attached.  A small secondary grain is outlined with a marker.  Later, acid etching of the ingot reduced the size and number of the secondary grains indicating that they are shallow. }  
\end{figure}

Cross-sections 1-2mm thick were taken from near the start of the crystal
growth and from the end for characterization.  One side of each was polished
and looked at optically and with x-ray fluorescence (XRF).  From the XRF
measurements, the sample was found to be single phase and the end of the
growth to be Mn rich.  The samples were then etched in a 25\% by volume
solution of nitric acid in water.  Optically, the start of the growth is a
single phase, single crystal while the end of the growth has large grains with
2nd phase along the grain boundaries.  Small pits were seen both optically and
by XRF.  The pits could be minimized by varying polishing techniques, but not
gotten rid of.  Fig.~\ref{fig:ingot} displays the as-grown crystal.

Only the body portion of the crystal growth were used for the experiments.  The ends of the growth were looked at as part of the characterization, but were not used because the end of the growth contained multiple grains and a second phase.  An additional examination of the body waws done to ensure that enough of the bodyt had been cut away as to remove those unwanted elements.  The small shallow grains that remained on one end of the body were avoided when cutting the sample to be measured.  As mentioned above, the XRF showed the body of the crystal growthy to be single phase.  The composition gradient is gradual and smooth, and there was no evidence of a Mn inhomogeneity seen in either the XRF or optical characterization.

\begin{figure}[b]
	\centering
	\includegraphics[width=\columnwidth]{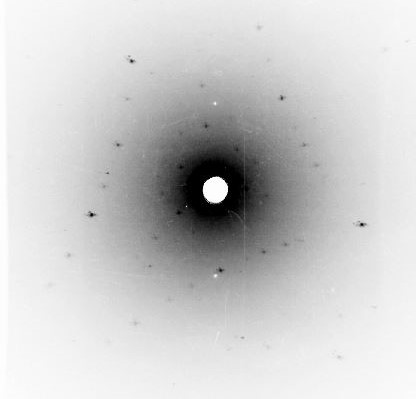}
	\caption{\label{fig:laue}Laue x-ray diffraction pattern of the sample confirms it is a single crystal, the $\text{Cu}_{94}\text{Mn}_{6}$ cube sample was etched in 15\% nitric acid.  The 6 at.\% Mn concentration was estimated from scaling using the observed temperature ($T_g=31.5$ K) at which the remanence disappeared.}
\end{figure}

Further investigation was done by polishing the cut ends of the ingot
body followed by etching.  No evidence of 2nd phase was seen and only
occasional small, shallow secondary grains were found.  In the
Bridgman method, it is not unusual for the very end of the growth to
be different because of accumulation of rejected elements and
impurities ahead of the growth front.  This would account for the
change in growth habit (increased number of grains), presence of 2nd
phase and overall Mn-rich composition seen at the end of the growth
but not in the body.  Laue x-ray diffraction along the length of the
body, Fig.~\ref{fig:laue}, confirms that the majority of the body is one single grain.

\section{The parameters for computing the Josephson length}\label{app:Josephson}

We follow here Ref.~\cite{janus:18}.  The first step is converting the
temperature to \emph{Janus units}
\begin{equation}
T^{(J)}=\frac{T}{\Tc}T_{\mathrm{c}}^{(J)}\,\ T_{\mathrm{c}}^{(J)}=1.102\,.
\end{equation}
Therefore, our working temperature $T=28K$ translates to $T^{(J)}=0.98$.

Next, we need to recall that the  only thing we know for
sure about this length scale is how it scales:
\begin{equation}
\ellJ(T^{(J)})=\frac{b_0+b_1(T_{\mathrm{c}}^{(J)}-T^{(J)})^\nu+b_2(T_{\mathrm{c}}^{(J)}-T^{(J)})^{\omega\nu}}{(T_{\mathrm{c}}^{(J)}-T^{(J)})^{-\nu}}
\end{equation}
where we include analytic ($b_1$) and confluent ($b_2)$ scaling
corrections with $\omega=1.12(10)$, $\nu=2.56(4)$ and
$T_{\mathrm{c}}^{(J)}=1.102(3)$~\cite{janus:13}. 
Now, although there is no unique way of fixing the overall scale $b_0$
(only the quotients $b_1/b_0$ and $b_2/b_0$ can be fixed in an unique
way), we shall adhere to the normalization of Ref.~\cite{janus:18}, so
that we can compare to their data in a direct way:
\begin{eqnarray}
b_0&=&0.101507196509469\,,\\
b_1&=&0.372545152960033\,,\\
b_2&=&0.199692833647175\,.
\end{eqnarray}
With this convention for $b_0$, at the working temperature $T=0.98$ we
have $\ellJ(0.98)=21.82\,a_0$.

\section{The replicon exponent and the self-consistent computation of $\xi$}\label{app:replicon}
Let us recall from the main text, the relation linking the number of
correlated spins $N_c$ with the correlation length $\xi$:
\begin{equation}\label{eqSM:Nc-xi}
N_c\approx \bigg({\frac {\xi}{a_0}}\bigg)^{d_f}\,,
\end{equation}
where $d_f$ is the fractal dimension equal to $d_f=d-\theta/2$ ($d=3$ is the
space dimension, while $\theta$ is the so-called replicon
exponent~\cite{janus:17b}). The quantity directly measured in the experiment
is $Nc$, and our goal is to convert it into a length by using the fractal
dimension.

\begin{figure}[t]
\includegraphics[width=\linewidth]{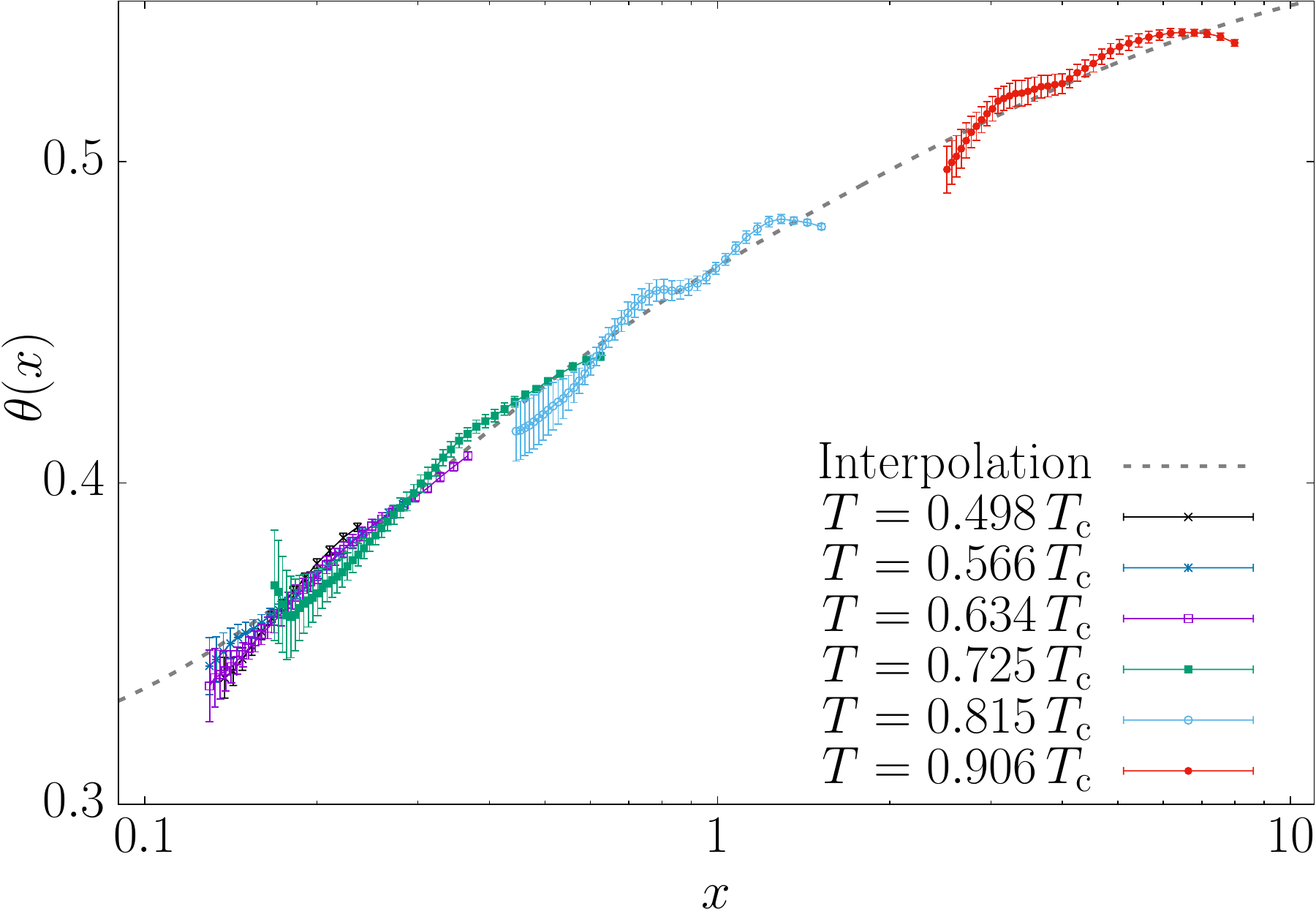}
\caption{Data for the replicon exponent, taken from Fig. 5 in the
  Supplemental Material for Ref.~\cite{janus:18}, as a function of the
  crossover variable $x$ defined in Eq.~\eqref{eqSM:x-def}. The black
  line is the RSB-inspired interpolation in
  Eq.~\eqref{eqSM:RSB-interpolation}. The wiggles are due to the extreme
  data-correlation (see, e.g, the discussion of Fig.~1 in
  Ref.~\cite{janus:08b}).
\label{figSM:interpolation}}
\end{figure}
Now, the problem with Eq.~\eqref{eqSM:Nc-xi} is that
the replicon exponent, and hence $d_f$, depends on both the temperature and
$\xi$ through the crossover variable (for the reader's convenience, we repeat
here thee definitions given in the main text):
\begin{equation}\label{eqSM:x-def}
x=\frac{\ellJ(T)}{\xi(\tw,T)}\,.
\end{equation}
The data for $\theta(x(\xi,T))$, as well as a discussion of the asymptotic
behavior for small $x$, are given in Sect.~C of the Supplemental Material (SM)
for~\cite{janus:18}. Here, we only observe that the numerical data for
$\theta(x(\xi,T))$ are well interpolated as (see Fig.~\ref{figSM:interpolation})
\begin{equation}\label{eqSM:RSB-interpolation}
\theta(x)=\theta_0 + d_1\Big(\frac{x}{1+e_1x}\Big)^{2-\theta_0}+
d_2\Big(\frac{x}{1+e_2x}\Big)^{3-\theta_0}\,,
\end{equation}
with numerical coefficients
\begin{eqnarray}
\theta_0\          &=& 0.303980\,,\\
e_1              &=& 1.38179\,,\\
d_1              &=& 2.72489\,,\\
e_2              &=& 2.12634\,,\\
d_2              &=& -9.98359\,.
\end{eqnarray}
Let us emphasize that the interpolation~\ref{eqSM:RSB-interpolation} is
consistent with the Replica Symmetry Breaking (RSB) asymptotic analysis (for
small $x$) presented in the SM for~\cite{janus:18}. Yet,
Eq.~\eqref{eqSM:RSB-interpolation} can be applied as well for larger $x$ if
needed.

Now, a droplets model supporter will object that $\theta_0$ should be zero
(according to their theory). However, the RSB/droplets controversy is
immaterial here: data can be fitted as well to the droplet model
(see~\cite{janus:18}), but the droplets fit start to depart significantly from
the RSB interpolation in Eq.~\eqref{eqSM:RSB-interpolation} only for
$x<0.065$. Because we aim to use the interpolation in the range $x\geq
0.0915$, we do not need to care about the RSB/droplets controversy.

\begin{figure}[b]
\includegraphics[width=\linewidth]{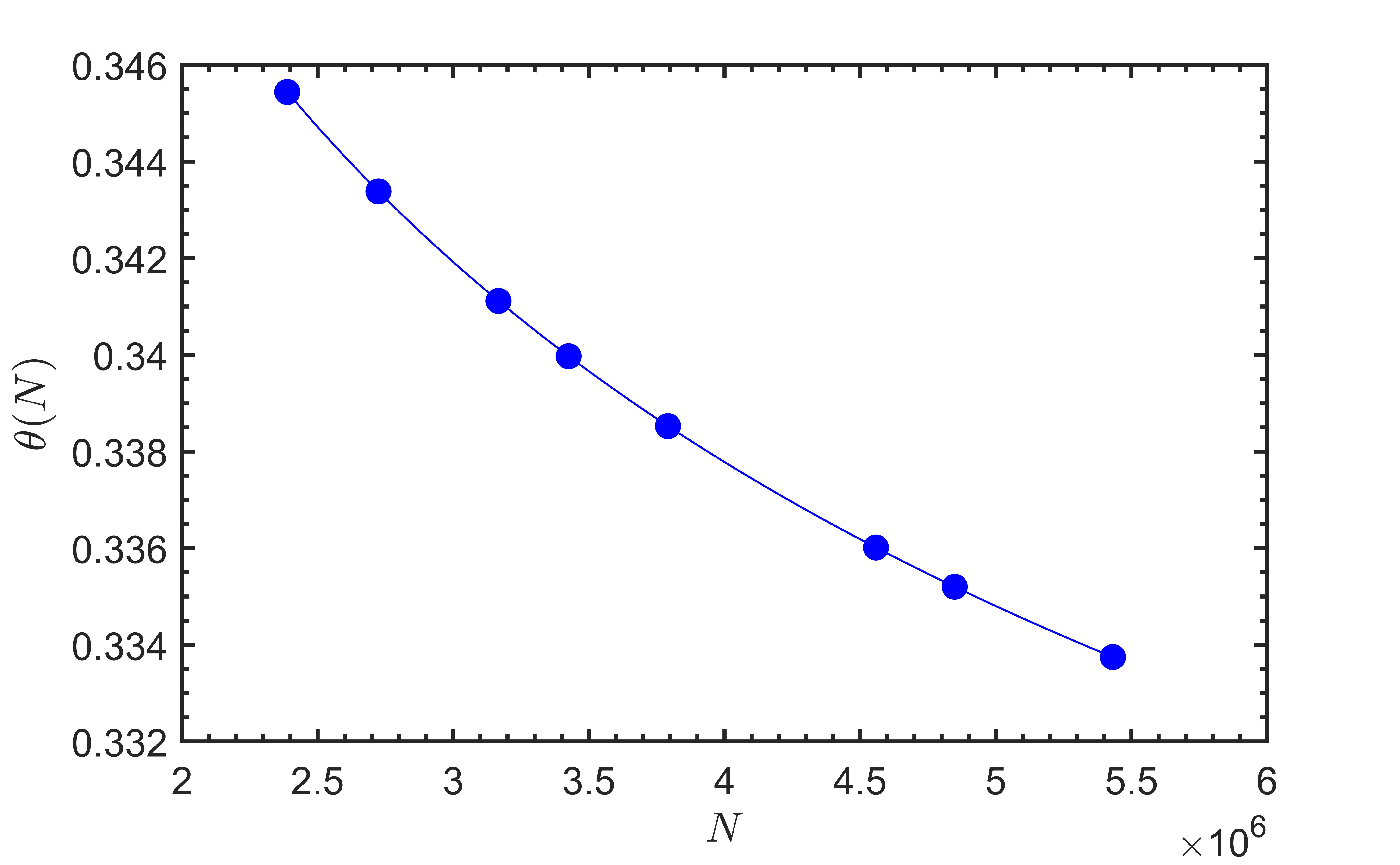}
\caption{Self consistent computation of the replicon exponent $\theta$. By
  varying $\xi$, we obtain a parametric plot of
  $\theta=\theta(x=\ellJ(T=28K)/\xi)$, Eq.~\eqref{eqSM:RSB-interpolation},
  as a function of the measured number of correlated spins $N_c$, see
  Eq.~\eqref{eqSM:parametric}. The dots are the appropriate values of $\theta$
  for our measured $N_c$. Note that $\theta$ is essentially constant in the
  experimentally relevant range of $Nc$.
\label{figSM:self-consistent}}
\end{figure}
After these preliminaries, the self-consistent computation is straightforward.
In order to obtain $\theta$ as a function of the measured number of correlated
spins $N_c$, see Fig.~\ref{figSM:self-consistent}, we just need to vary $\xi$
parametrically and compute both $\theta(x=\ellJ(T=28K)/\xi)$ from
Eq.~\eqref{eqSM:RSB-interpolation} and $N_c$ from
\begin{equation}\label{eqSM:parametric}
N_c=(\xi/a_0)^{d_f(x=\ellJ(28K)/\xi)}\,,d_f(x)=3-\frac{\theta{(x)}}{2}\,.
\end{equation}

\section{Details on the extrapolation of the aging rate}\label{app:extrapolation}

Our basic quantity will be the (bare) aging-rate
\begin{equation}
z(T,\xi)=\frac{\mathrm{d}\, \ln\, \tw}{\mathrm{d}\, \ln\, \xi}\,
\end{equation}
(the renormalized aging-rate considered in the main text is just
$z_{\text{c}}=z T/\Tc$).

Our starting point will be Table III in the Supplemental Material for
Ref.~\cite{janus:18}. In this table, we find the extrapolated bare aging-rates
for $\xi_{\text{target}}=38\, a_0, 76\, a_0$ and $\infty$, as computed from
the convergent ansatz:
\begin{equation}\label{eqSM:convergent-ansatz}
z(T,\xi_{\text{target}},\xi_{\text{min}})=z_\infty(T,\xi_{\text{min}})+\frac{A(T,\xi_{\text{min}})}{\xi_{\text{target}}^{\hat\omega}}\,.
\end{equation}
In the above expression, $\hat\omega=0.35$ and $\xi_\mathrm{min}$ is the
minimal correlation-length considered in their fit. It varies from varies from
$\xi_\mathrm{min}=3.5\, a_0$ to $\xi_\mathrm{min}=9\, a_0$ (or less than
$9\,a_0$ at the lowest temperatures).

Our first step was getting the slopes $A(T,\xi_{\text{min}})$ from the
tabulated values for $\xi_{\text{target}}=38\, a_0$ and $76\, a_0$ (instead,
$z_\infty(T,\xi_{\text{min}})$ is directly tabulated). With this information
in our hands, we may compute $z(T,\xi_{\text{target}},\xi_{\text{min}})$ for
any value of $\xi_{\text{target}}$ we wish. As for the error estimate, it is
only slightly more complicated:
\begin{eqnarray}
\Delta^2 z(T,\xi_{\text{target}},\xi_\text{min})&=&
E_{11}^{(T,\xi_\text{min})} + E_{22}^{(T,\xi_\text{min})}\frac
{1}{\xi_{\text{target}}^{2\hat\omega}}\nonumber\\
&+& E_{12}^{(T,\xi_\text{min})}\frac {1}{\xi_{\text{target}}^{\hat\omega}}\,.\label{eqSM:errors_convergent-ansatz}
\end{eqnarray}
Now, for every $T$ and $\xi_\text{min}$, we find error estimates for
$\xi_{\text{target}}=\infty$, $76\, a_0$ and $38\, a_0$ in the table by the
Janus Collaboration, which allows us to obtain the constants
$E_{11}^{(T,\xi_\text{min})}$, $E_{22}^{(T,\xi_\text{min})}$.  Once we have in
our hands the coefficients $E_{22}^{(T,\xi_\text{min})}$,
$E_{11}^{(T,\xi_\text{min})}$ and $E_{12}^{(T,\xi,\xi_\text{min})}$ we may
compute errors for whatever value of $\xi_{\text{target}}$ we need by using
Eq~\eqref{eqSM:errors_convergent-ansatz}.

Our next step was obtaining $z(T,\xi_{\text{target}},\xi_\text{min})$ for
a grid of values $180.26\,a_0 \leq \xi_{\text{target}}\leq 238.34\, a_0$.  We
computed $z(T,\xi_{\text{target}},\xi_\text{min})$ for all the values of
$(T,\xi_\text{min})$ in their Table III. We only neglected the few entries
where the error for $z(T,\xi_{\text{target}}= \infty,\xi_\text{min})$ was well
above $10\%$. Then, the estimates for the different $(T,\xi_\text{min})$
but the same $\xi_{\text{target}}$ were combined as explained, in the main
text (recall that the renormalized aging rate is $\zc=Tz/\Tc$) by means
to a fit to:
\begin{equation}\label{eqSM:well-organized-data}
\zc(T,\xi_{\text{target}},\xi_{\text{min}})=6.69+\frac{\alpha(\xi_{\text{target}})}{x_{\text{min}}^{\beta(\xi_{\text{target}})}}\,\,,
\end{equation}
where $x_{\text{min}}=\ellJ(T)/\xi_{\text{min}}$. Our final extrapolation
was
\begin{equation}\label{eqSM:zc-final}
  \zc(\xi_\text{target})=6.69+
  \frac{\alpha(\xi_{\text{target}})}{x_{\text{target}}^{\beta(\xi_{\text{target}})}}\,,\ 
  x_{\text{target}}=\frac{\ellJ(28\text{K})}{\xi_{\text{target}}}\,.
\end{equation}
The only tricky point needing further discussion regards the computation of
errors in $\zc(\xi_{\text{target}})$. It is clear that the different data in
the fit are extremely correlated (at least those at the same temperature: in
Table III of the SM for Ref.~\cite{janus:18} the Janus collaboration was
simply using the same set of $\xi(\tw,T)$ and discarding those with
$\xi(\tw,T)<\xi_\text{min}$). Under such conditions, the fit's standard errors
are not reliable. Hence, in order to estimate errors, we simply repeated the
fit for $\zc(T,\xi_{\text{target}},\xi_{\text{min}})$ plus (or minus) the
error. In other words, we assumed coherent fluctuations for all the data
set. The errors quoted in the main text are the halved difference between the
fit with \emph{data plus error} and \emph{data minus error}. A second, far
more conservative error estimate, would be just taking the error from the data
point at the lowest value of $x_\text{min}$ included in the fit to
Eq.~\eqref{eqSM:well-organized-data}. The conservative error estimate is
larger than the error from the \emph{halved-difference} by a factor 3.75.

\bibliographystyle{apsrev4-1}
%\bibliography{biblio}

\begin{thebibliography}{30}%
\makeatletter
\providecommand \@ifxundefined [1]{%
 \@ifx{#1\undefined}
}%
\providecommand \@ifnum [1]{%
 \ifnum #1\expandafter \@firstoftwo
 \else \expandafter \@secondoftwo
 \fi
}%
\providecommand \@ifx [1]{%
 \ifx #1\expandafter \@firstoftwo
 \else \expandafter \@secondoftwo
 \fi
}%
\providecommand \natexlab [1]{#1}%
\providecommand \enquote  [1]{``#1''}%
\providecommand \bibnamefont  [1]{#1}%
\providecommand \bibfnamefont [1]{#1}%
\providecommand \citenamefont [1]{#1}%
\providecommand \href@noop [0]{\@secondoftwo}%
\providecommand \href [0]{\begingroup \@sanitize@url \@href}%
\providecommand \@href[1]{\@@startlink{#1}\@@href}%
\providecommand \@@href[1]{\endgroup#1\@@endlink}%
\providecommand \@sanitize@url [0]{\catcode `\\12\catcode `\$12\catcode
  `\&12\catcode `\#12\catcode `\^12\catcode `\_12\catcode `\%12\relax}%
\providecommand \@@startlink[1]{}%
\providecommand \@@endlink[0]{}%
\providecommand \url  [0]{\begingroup\@sanitize@url \@url }%
\providecommand \@url [1]{\endgroup\@href {#1}{\urlprefix }}%
\providecommand \urlprefix  [0]{URL }%
\providecommand \Eprint [0]{\href }%
\providecommand \doibase [0]{http://dx.doi.org/}%
\providecommand \selectlanguage [0]{\@gobble}%
\providecommand \bibinfo  [0]{\@secondoftwo}%
\providecommand \bibfield  [0]{\@secondoftwo}%
\providecommand \translation [1]{[#1]}%
\providecommand \BibitemOpen [0]{}%
\providecommand \bibitemStop [0]{}%
\providecommand \bibitemNoStop [0]{.\EOS\space}%
\providecommand \EOS [0]{\spacefactor3000\relax}%
\providecommand \BibitemShut  [1]{\csname bibitem#1\endcsname}%
\let\auto@bib@innerbib\@empty
%</preamble>
\bibitem [{\citenamefont {Cavagna}(2009)}]{cavagna:09}%
  \BibitemOpen
  \bibfield  {author} {\bibinfo {author} {\bibfnamefont {A.}~\bibnamefont
  {Cavagna}},\ }\href {\doibase 10.1016/j.physrep.2009.03.003} {\bibfield
  {journal} {\bibinfo  {journal} {Physics Reports}\ }\textbf {\bibinfo {volume}
  {476}},\ \bibinfo {pages} {51} (\bibinfo {year} {2009})} \BibitemShut
  {NoStop}%
\bibitem [{\citenamefont {Gunnarsson}\ \emph {et~al.}(1991)\citenamefont
  {Gunnarsson}, \citenamefont {Svedlindh}, \citenamefont {Nordblad},
  \citenamefont {Lundgren}, \citenamefont {Aruga},\ and\ \citenamefont
  {Ito}}]{gunnarsson:91}%
  \BibitemOpen
  \bibfield  {author} {\bibinfo {author} {\bibfnamefont {K.}~\bibnamefont
  {Gunnarsson}}, \bibinfo {author} {\bibfnamefont {P.}~\bibnamefont
  {Svedlindh}}, \bibinfo {author} {\bibfnamefont {P.}~\bibnamefont {Nordblad}},
  \bibinfo {author} {\bibfnamefont {L.}~\bibnamefont {Lundgren}}, \bibinfo
  {author} {\bibfnamefont {H.}~\bibnamefont {Aruga}}, \ and\ \bibinfo {author}
  {\bibfnamefont {A.}~\bibnamefont {Ito}},\ }\href {\doibase
  10.1103/PhysRevB.43.8199} {\bibfield  {journal} {\bibinfo  {journal} {Phys.
  Rev. B}\ }\textbf {\bibinfo {volume} {43}},\ \bibinfo {pages} {8199}
  (\bibinfo {year} {1991})}\BibitemShut {NoStop}%
\bibitem [{\citenamefont {Adam}\ and\ \citenamefont {Gibbs}(1965)}]{adam:65}%
  \BibitemOpen
  \bibfield  {author} {\bibinfo {author} {\bibfnamefont {G.}~\bibnamefont
  {Adam}}\ and\ \bibinfo {author} {\bibfnamefont {J.~H.}\ \bibnamefont
  {Gibbs}},\ }\href {\doibase http://dx.doi.org/10.1063/1.1696442} {\bibfield
  {journal} {\bibinfo  {journal} {The Journal of Chemical Physics}\ }\textbf
  {\bibinfo {volume} {43}},\ \bibinfo {pages} {139} (\bibinfo {year}
  {1965})}\BibitemShut {NoStop}%
\bibitem [{\citenamefont {Joh}\ \emph {et~al.}(1999)\citenamefont {Joh},
  \citenamefont {Orbach}, \citenamefont {Wood}, \citenamefont {Hammann},\ and\
  \citenamefont {Vincent}}]{joh:99}%
  \BibitemOpen
  \bibfield  {author} {\bibinfo {author} {\bibfnamefont {Y.~G.}\ \bibnamefont
  {Joh}}, \bibinfo {author} {\bibfnamefont {R.}~\bibnamefont {Orbach}},
  \bibinfo {author} {\bibfnamefont {G.~G.}\ \bibnamefont {Wood}}, \bibinfo
  {author} {\bibfnamefont {J.}~\bibnamefont {Hammann}}, \ and\ \bibinfo
  {author} {\bibfnamefont {E.}~\bibnamefont {Vincent}},\ }\href {\doibase
  10.1103/PhysRevLett.82.438} {\bibfield  {journal} {\bibinfo  {journal} {Phys.
  Rev. Lett.}\ }\textbf {\bibinfo {volume} {82}},\ \bibinfo {pages} {438}
  (\bibinfo {year} {1999})}\BibitemShut {NoStop}%
\bibitem [{\citenamefont {Albert}\ \emph {et~al.}(2016)\citenamefont {Albert},
  \citenamefont {Bauer}, \citenamefont {Michl}, \citenamefont {Biroli},
  \citenamefont {Bouchaud}, \citenamefont {Loidl}, \citenamefont
  {Lunkenheimer}, \citenamefont {Tourbot}, \citenamefont {Wiertel-Gasquet},\
  and\ \citenamefont {Ladieu}}]{albert:16}%
  \BibitemOpen
  \bibfield  {author} {\bibinfo {author} {\bibfnamefont {S.}~\bibnamefont
  {Albert}}, \bibinfo {author} {\bibfnamefont {T.}~\bibnamefont {Bauer}},
  \bibinfo {author} {\bibfnamefont {M.}~\bibnamefont {Michl}}, \bibinfo
  {author} {\bibfnamefont {G.}~\bibnamefont {Biroli}}, \bibinfo {author}
  {\bibfnamefont {J.-P.}\ \bibnamefont {Bouchaud}}, \bibinfo {author}
  {\bibfnamefont {A.}~\bibnamefont {Loidl}}, \bibinfo {author} {\bibfnamefont
  {P.}~\bibnamefont {Lunkenheimer}}, \bibinfo {author} {\bibfnamefont
  {R.}~\bibnamefont {Tourbot}}, \bibinfo {author} {\bibfnamefont
  {C.}~\bibnamefont {Wiertel-Gasquet}}, \ and\ \bibinfo {author} {\bibfnamefont
  {F.}~\bibnamefont {Ladieu}},\ }\href {\doibase 10.1126/science.aaf3182}
  {\bibfield  {journal} {\bibinfo  {journal} {Science}\ }\textbf {\bibinfo
  {volume} {352}},\ \bibinfo {pages} {1308} (\bibinfo {year} {2016})} \BibitemShut
  {NoStop}%
\bibitem [{Note1()}]{Note1}%
  \BibitemOpen
  \bibinfo {note} {General theoretical arguments suggest that $\protect
  \ensuremath {z_\protect \mathrm {c}}\protect \xspace $ is also $\xi
  $-independent at exactly $T=\protect \ensuremath {T_\protect \mathrm
  {c}}\protect \xspace $~\cite {zinn-justin:05}. Only if $\protect \ensuremath
  {z_\protect \mathrm {c}}\protect \xspace $ is $\xi $-independent we have a
  power-law scaling $\protect \ensuremath {t_\protect \mathrm {w}}\protect
  \xspace \propto \xi ^{\protect \ensuremath {T_\protect \mathrm {c}}\protect
  \xspace \protect \ensuremath {z_\protect \mathrm {c}}\protect \xspace /T}$.
  If $\protect \ensuremath {z_\protect \mathrm {c}}\protect \xspace $ grows
  with $\xi $, as we find here, we encounter a dynamics slower than a power-law
  (for instance, an activated dynamics with free-energy barriers $\Delta
  \propto \xi ^\Psi $ for some $\Psi >0$).}\BibitemShut {Stop}%
\bibitem [{\citenamefont {Zhai}\ \emph {et~al.}(2017)\citenamefont {Zhai},
  \citenamefont {Harrison}, \citenamefont {Tennant}, \citenamefont {Dalhberg},
  \citenamefont {Kenning},\ and\ \citenamefont {Orbach}}]{zhai:17}%
  \BibitemOpen
  \bibfield  {author} {\bibinfo {author} {\bibfnamefont {Q.}~\bibnamefont
  {Zhai}}, \bibinfo {author} {\bibfnamefont {D.~C.}\ \bibnamefont {Harrison}},
  \bibinfo {author} {\bibfnamefont {D.}~\bibnamefont {Tennant}}, \bibinfo
  {author} {\bibfnamefont {E.~D.}\ \bibnamefont {Dalhberg}}, \bibinfo {author}
  {\bibfnamefont {G.~G.}\ \bibnamefont {Kenning}}, \ and\ \bibinfo {author}
  {\bibfnamefont {R.~L.}\ \bibnamefont {Orbach}},\ }\href {\doibase
  10.1103/PhysRevB.95.054304} {\bibfield  {journal} {\bibinfo  {journal} {Phys.
  Rev. B}\ }\textbf {\bibinfo {volume} {95}},\ \bibinfo {pages} {054304}
  (\bibinfo {year} {2017})}\BibitemShut {NoStop}%
\bibitem [{\citenamefont {Kenning}\ \emph {et~al.}(2018)\citenamefont
  {Kenning}, \citenamefont {Tennant}, \citenamefont {Rost}, \citenamefont
  {da~Silva}, \citenamefont {Walters}, \citenamefont {Zhai}, \citenamefont
  {Harrison}, \citenamefont {Dahlberg},\ and\ \citenamefont
  {Orbach}}]{kenning:18}%
  \BibitemOpen
  \bibfield  {author} {\bibinfo {author} {\bibfnamefont {G.~G.}\ \bibnamefont
  {Kenning}}, \bibinfo {author} {\bibfnamefont {D.~M.}\ \bibnamefont
  {Tennant}}, \bibinfo {author} {\bibfnamefont {C.~M.}\ \bibnamefont {Rost}},
  \bibinfo {author} {\bibfnamefont {F.~G.}\ \bibnamefont {da~Silva}}, \bibinfo
  {author} {\bibfnamefont {B.~J.}\ \bibnamefont {Walters}}, \bibinfo {author}
  {\bibfnamefont {Q.}~\bibnamefont {Zhai}}, \bibinfo {author} {\bibfnamefont
  {D.~C.}\ \bibnamefont {Harrison}}, \bibinfo {author} {\bibfnamefont {E.~D.}\
  \bibnamefont {Dahlberg}}, \ and\ \bibinfo {author} {\bibfnamefont {R.~L.}\
  \bibnamefont {Orbach}},\ }\href {\doibase 10.1103/PhysRevB.98.104436}
  {\bibfield  {journal} {\bibinfo  {journal} {Phys. Rev. B}\ }\textbf {\bibinfo
  {volume} {98}},\ \bibinfo {pages} {104436} (\bibinfo {year}
  {2018})}\BibitemShut {NoStop}%
\bibitem [{\citenamefont {Baity-Jesi}\ \emph {et~al.}(2018)\citenamefont
  {Baity-Jesi}, \citenamefont {Calore}, \citenamefont {Cruz}, \citenamefont
  {Fernandez}, \citenamefont {Gil-Narvion}, \citenamefont {Gordillo-Guerrero},
  \citenamefont {I\~niguez}, \citenamefont {Maiorano}, \citenamefont
  {Marinari}, \citenamefont {Martin-Mayor}, \citenamefont {Moreno-Gordo},
  \citenamefont {Mu\~noz Sudupe}, \citenamefont {Navarro}, \citenamefont
  {Parisi}, \citenamefont {Perez-Gaviro}, \citenamefont {Ricci-Tersenghi},
  \citenamefont {Ruiz-Lorenzo}, \citenamefont {Schifano}, \citenamefont
  {Seoane}, \citenamefont {Tarancon}, \citenamefont {Tripiccione},\ and\
  \citenamefont {Yllanes}}]{janus:18}%
  \BibitemOpen
  \bibfield  {author} {\bibinfo {author} {\bibfnamefont {M.}~\bibnamefont
  {Baity-Jesi}}, \bibinfo {author} {\bibfnamefont {E.}~\bibnamefont {Calore}},
  \bibinfo {author} {\bibfnamefont {A.}~\bibnamefont {Cruz}}, \bibinfo {author}
  {\bibfnamefont {L.~A.}\ \bibnamefont {Fernandez}}, \bibinfo {author}
  {\bibfnamefont {J.~M.}\ \bibnamefont {Gil-Narvion}}, \bibinfo {author}
  {\bibfnamefont {A.}~\bibnamefont {Gordillo-Guerrero}}, \bibinfo {author}
  {\bibfnamefont {D.}~\bibnamefont {I\~niguez}}, \bibinfo {author}
  {\bibfnamefont {A.}~\bibnamefont {Maiorano}}, \bibinfo {author}
  {\bibfnamefont {E.}~\bibnamefont {Marinari}}, \bibinfo {author}
  {\bibfnamefont {V.}~\bibnamefont {Martin-Mayor}}, \bibinfo {author}
  {\bibfnamefont {J.}~\bibnamefont {Moreno-Gordo}}, \bibinfo {author}
  {\bibfnamefont {A.}~\bibnamefont {Mu\~noz Sudupe}}, \bibinfo {author}
  {\bibfnamefont {D.}~\bibnamefont {Navarro}}, \bibinfo {author} {\bibfnamefont
  {G.}~\bibnamefont {Parisi}}, \bibinfo {author} {\bibfnamefont
  {S.}~\bibnamefont {Perez-Gaviro}}, \bibinfo {author} {\bibfnamefont
  {F.}~\bibnamefont {Ricci-Tersenghi}}, \bibinfo {author} {\bibfnamefont
  {J.~J.}\ \bibnamefont {Ruiz-Lorenzo}}, \bibinfo {author} {\bibfnamefont
  {S.~F.}\ \bibnamefont {Schifano}}, \bibinfo {author} {\bibfnamefont
  {B.}~\bibnamefont {Seoane}}, \bibinfo {author} {\bibfnamefont
  {A.}~\bibnamefont {Tarancon}}, \bibinfo {author} {\bibfnamefont
  {R.}~\bibnamefont {Tripiccione}}, \ and\ \bibinfo {author} {\bibfnamefont
  {D.}~\bibnamefont {Yllanes}} (\bibinfo {collaboration} {Janus
  Collaboration}),\ }\href {\doibase 10.1103/PhysRevLett.120.267203} {\bibfield
   {journal} {\bibinfo  {journal} {Phys. Rev. Lett.}\ }\textbf {\bibinfo
  {volume} {120}},\ \bibinfo {pages} {267203} (\bibinfo {year}
  {2018})}\BibitemShut {NoStop}%
\bibitem [{\citenamefont {Baity-Jesi}\ \emph
  {et~al.}(2014{\natexlab{a}})\citenamefont {Baity-Jesi}, \citenamefont
  {Ba\~{n}os}, \citenamefont {Cruz}, \citenamefont {Fernandez}, \citenamefont
  {Gil-Narvion}, \citenamefont {Gordillo-Guerrero}, \citenamefont {Iniguez},
  \citenamefont {Maiorano}, \citenamefont {Mantovani}, \citenamefont
  {Marinari}, \citenamefont {Mart\'{i}n-Mayor}, \citenamefont
  {Monforte-Garcia}, \citenamefont {Mu{\~n}oz~Sudupe}, \citenamefont {Navarro},
  \citenamefont {Parisi}, \citenamefont {Perez-Gaviro}, \citenamefont
  {Pivanti}, \citenamefont {Ricci-Tersenghi}, \citenamefont {Ruiz-Lorenzo},
  \citenamefont {Schifano}, \citenamefont {Seoane}, \citenamefont {Tarancon},
  \citenamefont {Tripiccione},\ and\ \citenamefont {Yllanes}}]{janus:14}%
  \BibitemOpen
  \bibfield  {author} {\bibinfo {author} {\bibfnamefont {M.}~\bibnamefont
  {Baity-Jesi}}, \bibinfo {author} {\bibfnamefont {R.~A.}\ \bibnamefont
  {Ba\~{n}os}}, \bibinfo {author} {\bibfnamefont {A.}~\bibnamefont {Cruz}},
  \bibinfo {author} {\bibfnamefont {L.~A.}\ \bibnamefont {Fernandez}}, \bibinfo
  {author} {\bibfnamefont {J.~M.}\ \bibnamefont {Gil-Narvion}}, \bibinfo
  {author} {\bibfnamefont {A.}~\bibnamefont {Gordillo-Guerrero}}, \bibinfo
  {author} {\bibfnamefont {D.}~\bibnamefont {Iniguez}}, \bibinfo {author}
  {\bibfnamefont {A.}~\bibnamefont {Maiorano}}, \bibinfo {author}
  {\bibfnamefont {F.}~\bibnamefont {Mantovani}}, \bibinfo {author}
  {\bibfnamefont {E.}~\bibnamefont {Marinari}}, \bibinfo {author}
  {\bibfnamefont {V.}~\bibnamefont {Mart\'{i}n-Mayor}}, \bibinfo {author}
  {\bibfnamefont {J.}~\bibnamefont {Monforte-Garcia}}, \bibinfo {author}
  {\bibfnamefont {A.}~\bibnamefont {Mu{\~n}oz~Sudupe}}, \bibinfo {author}
  {\bibfnamefont {D.}~\bibnamefont {Navarro}}, \bibinfo {author} {\bibfnamefont
  {G.}~\bibnamefont {Parisi}}, \bibinfo {author} {\bibfnamefont
  {S.}~\bibnamefont {Perez-Gaviro}}, \bibinfo {author} {\bibfnamefont
  {M.}~\bibnamefont {Pivanti}}, \bibinfo {author} {\bibfnamefont
  {F.}~\bibnamefont {Ricci-Tersenghi}}, \bibinfo {author} {\bibfnamefont
  {J.~J.}\ \bibnamefont {Ruiz-Lorenzo}}, \bibinfo {author} {\bibfnamefont
  {S.~F.}\ \bibnamefont {Schifano}}, \bibinfo {author} {\bibfnamefont
  {B.}~\bibnamefont {Seoane}}, \bibinfo {author} {\bibfnamefont
  {A.}~\bibnamefont {Tarancon}}, \bibinfo {author} {\bibfnamefont
  {R.}~\bibnamefont {Tripiccione}}, \ and\ \bibinfo {author} {\bibfnamefont
  {D.}~\bibnamefont {Yllanes}} (\bibinfo {collaboration} {Janus
  Collaboration}),\ }\href {\doibase 10.1016/j.cpc.2013.10.019} {\bibfield
  {journal} {\bibinfo  {journal} {Comp. Phys. Comm}\ }\textbf {\bibinfo
  {volume} {185}},\ \bibinfo {pages} {550} (\bibinfo {year}
  {2014}{\natexlab{a}})} \BibitemShut {NoStop}%
\bibitem [{\citenamefont {Josephson}(1966)}]{josephson:66}%
  \BibitemOpen
  \bibfield  {author} {\bibinfo {author} {\bibfnamefont {B.~D.}\ \bibnamefont
  {Josephson}},\ }\href@noop {} {\bibfield  {journal} {\bibinfo  {journal}
  {Phys. Lett.}\ }\textbf {\bibinfo {volume} {21}},\ \bibinfo {pages} {608}
  (\bibinfo {year} {1966})}\BibitemShut {NoStop}%
\bibitem [{\citenamefont {Baity-Jesi}\ \emph {et~al.}(2013)\citenamefont
  {Baity-Jesi}, \citenamefont {Ba\~{n}os}, \citenamefont {Cruz}, \citenamefont
  {Fernandez}, \citenamefont {Gil-Narvion}, \citenamefont {Gordillo-Guerrero},
  \citenamefont {Iniguez}, \citenamefont {Maiorano}, \citenamefont {Mantovani},
  \citenamefont {Marinari}, \citenamefont {Mart\'{i}n-Mayor}, \citenamefont
  {Monforte-Garcia}, \citenamefont {Mu{\~n}oz~Sudupe}, \citenamefont {Navarro},
  \citenamefont {Parisi}, \citenamefont {Perez-Gaviro}, \citenamefont
  {Pivanti}, \citenamefont {Ricci-Tersenghi}, \citenamefont {Ruiz-Lorenzo},
  \citenamefont {Schifano}, \citenamefont {Seoane}, \citenamefont {Tarancon},
  \citenamefont {Tripiccione},\ and\ \citenamefont {Yllanes}}]{janus:13}%
  \BibitemOpen
  \bibfield  {author} {\bibinfo {author} {\bibfnamefont {M.}~\bibnamefont
  {Baity-Jesi}}, \bibinfo {author} {\bibfnamefont {R.~A.}\ \bibnamefont
  {Ba\~{n}os}}, \bibinfo {author} {\bibfnamefont {A.}~\bibnamefont {Cruz}},
  \bibinfo {author} {\bibfnamefont {L.~A.}\ \bibnamefont {Fernandez}}, \bibinfo
  {author} {\bibfnamefont {J.~M.}\ \bibnamefont {Gil-Narvion}}, \bibinfo
  {author} {\bibfnamefont {A.}~\bibnamefont {Gordillo-Guerrero}}, \bibinfo
  {author} {\bibfnamefont {D.}~\bibnamefont {Iniguez}}, \bibinfo {author}
  {\bibfnamefont {A.}~\bibnamefont {Maiorano}}, \bibinfo {author}
  {\bibfnamefont {F.}~\bibnamefont {Mantovani}}, \bibinfo {author}
  {\bibfnamefont {E.}~\bibnamefont {Marinari}}, \bibinfo {author}
  {\bibfnamefont {V.}~\bibnamefont {Mart\'{i}n-Mayor}}, \bibinfo {author}
  {\bibfnamefont {J.}~\bibnamefont {Monforte-Garcia}}, \bibinfo {author}
  {\bibfnamefont {A.}~\bibnamefont {Mu{\~n}oz~Sudupe}}, \bibinfo {author}
  {\bibfnamefont {D.}~\bibnamefont {Navarro}}, \bibinfo {author} {\bibfnamefont
  {G.}~\bibnamefont {Parisi}}, \bibinfo {author} {\bibfnamefont
  {S.}~\bibnamefont {Perez-Gaviro}}, \bibinfo {author} {\bibfnamefont
  {M.}~\bibnamefont {Pivanti}}, \bibinfo {author} {\bibfnamefont
  {F.}~\bibnamefont {Ricci-Tersenghi}}, \bibinfo {author} {\bibfnamefont
  {J.~J.}\ \bibnamefont {Ruiz-Lorenzo}}, \bibinfo {author} {\bibfnamefont
  {S.~F.}\ \bibnamefont {Schifano}}, \bibinfo {author} {\bibfnamefont
  {B.}~\bibnamefont {Seoane}}, \bibinfo {author} {\bibfnamefont
  {A.}~\bibnamefont {Tarancon}}, \bibinfo {author} {\bibfnamefont
  {R.}~\bibnamefont {Tripiccione}}, \ and\ \bibinfo {author} {\bibfnamefont
  {D.}~\bibnamefont {Yllanes}} (\bibinfo {collaboration} {Janus
  Collaboration}),\ }\href {\doibase 10.1103/PhysRevB.88.224416} {\bibfield
  {journal} {\bibinfo  {journal} {Phys. Rev. B}\ }\textbf {\bibinfo {volume}
  {88}},\ \bibinfo {pages} {224416} (\bibinfo {year} {{2013}})} \BibitemShut
  {NoStop}%
\bibitem [{\citenamefont {Bray}\ and\ \citenamefont {Moore}(1982)}]{bray:82}%
  \BibitemOpen
  \bibfield  {author} {\bibinfo {author} {\bibfnamefont {A.~J.}\ \bibnamefont
  {Bray}}\ and\ \bibinfo {author} {\bibfnamefont {M.~A.}\ \bibnamefont
  {Moore}},\ }\href {\doibase 10.1088/0022-3719/15/18/007} {\bibfield
  {journal} {\bibinfo  {journal} {J. Phys. C: Solid St. Phys.}\ }\textbf
  {\bibinfo {volume} {15}},\ \bibinfo {pages} {3897} (\bibinfo {year}
  {1982})}\BibitemShut {NoStop}%
\bibitem [{\citenamefont {Baity-Jesi}\ \emph
  {et~al.}(2014{\natexlab{b}})\citenamefont {Baity-Jesi}, \citenamefont
  {Fernandez}, \citenamefont {Mart\'{i}n-Mayor},\ and\ \citenamefont
  {Sanz}}]{baityjesi:14}%
  \BibitemOpen
  \bibfield  {author} {\bibinfo {author} {\bibfnamefont {M.}~\bibnamefont
  {Baity-Jesi}}, \bibinfo {author} {\bibfnamefont {L.~A.}\ \bibnamefont
  {Fernandez}}, \bibinfo {author} {\bibfnamefont {V.}~\bibnamefont
  {Mart\'{i}n-Mayor}}, \ and\ \bibinfo {author} {\bibfnamefont {J.~M.}\
  \bibnamefont {Sanz}},\ }\href {\doibase 10.1103/PhysRevB.89.014202}
  {\bibfield  {journal} {\bibinfo  {journal} {Phys. Rev.}\ }\textbf {\bibinfo
  {volume} {89}},\ \bibinfo {pages} {014202} (\bibinfo {year}
  {2014}{\natexlab{b}})} \BibitemShut {NoStop}%
\bibitem [{\citenamefont {Amit}\ and\ \citenamefont
  {Mart\'{i}n-Mayor}(2005)}]{amit:05}%
  \BibitemOpen
  \bibfield  {author} {\bibinfo {author} {\bibfnamefont {D.~J.}\ \bibnamefont
  {Amit}}\ and\ \bibinfo {author} {\bibfnamefont {V.}~\bibnamefont
  {Mart\'{i}n-Mayor}},\ }\href {\doibase 10.1142/9789812775313_bmatter} {\emph
  {\bibinfo {title} {Field Theory, the Renormalization Group and Critical
  Phenomena}}},\ \bibinfo {edition} {3rd}\ ed.\ (\bibinfo  {publisher} {World
  Scientific},\ \bibinfo {address} {Singapore},\ \bibinfo {year}
  {2005})\BibitemShut {NoStop}%
\bibitem [{\citenamefont {Guchhait}\ and\ \citenamefont
  {Orbach}(2017)}]{guchhait:17}%
  \BibitemOpen
  \bibfield  {author} {\bibinfo {author} {\bibfnamefont {S.}~\bibnamefont
  {Guchhait}}\ and\ \bibinfo {author} {\bibfnamefont {R.~L.}\ \bibnamefont
  {Orbach}},\ }\href {\doibase 10.1103/PhysRevLett.118.157203} {\bibfield
  {journal} {\bibinfo  {journal} {Phys. Rev. Lett.}\ }\textbf {\bibinfo
  {volume} {118}},\ \bibinfo {pages} {157203} (\bibinfo {year}
  {2017})}\BibitemShut {NoStop}%
\bibitem [{\citenamefont {Wandersman}\ \emph {et~al.}(2008)\citenamefont
  {Wandersman}, \citenamefont {Dupuis}, \citenamefont {Dubois}, \citenamefont
  {Perzynski}, \citenamefont {Nakamae},\ and\ \citenamefont
  {Vincent}}]{Wandersman_2008}%
  \BibitemOpen
  \bibfield  {author} {\bibinfo {author} {\bibfnamefont {E.}~\bibnamefont
  {Wandersman}}, \bibinfo {author} {\bibfnamefont {V.}~\bibnamefont {Dupuis}},
  \bibinfo {author} {\bibfnamefont {E.}~\bibnamefont {Dubois}}, \bibinfo
  {author} {\bibfnamefont {R.}~\bibnamefont {Perzynski}}, \bibinfo {author}
  {\bibfnamefont {S.}~\bibnamefont {Nakamae}}, \ and\ \bibinfo {author}
  {\bibfnamefont {E.}~\bibnamefont {Vincent}},\ }\href {\doibase
  10.1209/0295-5075/84/37011} {\bibfield  {journal} {\bibinfo  {journal} {{EPL}
  (Europhysics Letters)}\ }\textbf {\bibinfo {volume} {84}},\ \bibinfo {pages}
  {37011} (\bibinfo {year} {2008})}\BibitemShut {NoStop}%
\bibitem [{\citenamefont {Nair}\ and\ \citenamefont
  {Nigam}(2007)}]{PhysRevB.75.214415}%
  \BibitemOpen
  \bibfield  {author} {\bibinfo {author} {\bibfnamefont {S.}~\bibnamefont
  {Nair}}\ and\ \bibinfo {author} {\bibfnamefont {A.~K.}\ \bibnamefont
  {Nigam}},\ }\href {\doibase 10.1103/PhysRevB.75.214415} {\bibfield  {journal}
  {\bibinfo  {journal} {Phys. Rev. B}\ }\textbf {\bibinfo {volume} {75}},\
  \bibinfo {pages} {214415} (\bibinfo {year} {2007})}\BibitemShut {NoStop}%
\bibitem [{\citenamefont {Baity-Jesi}\ \emph {et~al.}(2017)\citenamefont
  {Baity-Jesi}, \citenamefont {Calore}, \citenamefont {Cruz}, \citenamefont
  {Fernandez}, \citenamefont {Gil-Narvion}, \citenamefont {Gordillo-Guerrero},
  \citenamefont {I\~niguez}, \citenamefont {Maiorano}, \citenamefont
  {Marinari}, \citenamefont {Martin-Mayor}, \citenamefont {Monforte-Garcia},
  \citenamefont {Mu\~noz Sudupe}, \citenamefont {Navarro}, \citenamefont
  {Parisi}, \citenamefont {Perez-Gaviro}, \citenamefont {Ricci-Tersenghi},
  \citenamefont {Ruiz-Lorenzo}, \citenamefont {Schifano}, \citenamefont
  {Seoane}, \citenamefont {Tarancon}, \citenamefont {Tripiccione},\ and\
  \citenamefont {Yllanes}}]{janus:17b}%
  \BibitemOpen
  \bibfield  {author} {\bibinfo {author} {\bibfnamefont {M.}~\bibnamefont
  {Baity-Jesi}}, \bibinfo {author} {\bibfnamefont {E.}~\bibnamefont {Calore}},
  \bibinfo {author} {\bibfnamefont {A.}~\bibnamefont {Cruz}}, \bibinfo {author}
  {\bibfnamefont {L.~A.}\ \bibnamefont {Fernandez}}, \bibinfo {author}
  {\bibfnamefont {J.~M.}\ \bibnamefont {Gil-Narvion}}, \bibinfo {author}
  {\bibfnamefont {A.}~\bibnamefont {Gordillo-Guerrero}}, \bibinfo {author}
  {\bibfnamefont {D.}~\bibnamefont {I\~niguez}}, \bibinfo {author}
  {\bibfnamefont {A.}~\bibnamefont {Maiorano}}, \bibinfo {author}
  {\bibfnamefont {E.}~\bibnamefont {Marinari}}, \bibinfo {author}
  {\bibfnamefont {V.}~\bibnamefont {Martin-Mayor}}, \bibinfo {author}
  {\bibfnamefont {J.}~\bibnamefont {Monforte-Garcia}}, \bibinfo {author}
  {\bibfnamefont {A.}~\bibnamefont {Mu\~noz Sudupe}}, \bibinfo {author}
  {\bibfnamefont {D.}~\bibnamefont {Navarro}}, \bibinfo {author} {\bibfnamefont
  {G.}~\bibnamefont {Parisi}}, \bibinfo {author} {\bibfnamefont
  {S.}~\bibnamefont {Perez-Gaviro}}, \bibinfo {author} {\bibfnamefont
  {F.}~\bibnamefont {Ricci-Tersenghi}}, \bibinfo {author} {\bibfnamefont
  {J.~J.}\ \bibnamefont {Ruiz-Lorenzo}}, \bibinfo {author} {\bibfnamefont
  {S.~F.}\ \bibnamefont {Schifano}}, \bibinfo {author} {\bibfnamefont
  {B.}~\bibnamefont {Seoane}}, \bibinfo {author} {\bibfnamefont
  {A.}~\bibnamefont {Tarancon}}, \bibinfo {author} {\bibfnamefont
  {R.}~\bibnamefont {Tripiccione}}, \ and\ \bibinfo {author} {\bibfnamefont
  {D.}~\bibnamefont {Yllanes}} (\bibinfo {collaboration} {Janus
  Collaboration}),\ }\href {\doibase 10.1103/PhysRevLett.118.157202} {\bibfield
   {journal} {\bibinfo  {journal} {Phys. Rev. Lett.}\ }\textbf {\bibinfo
  {volume} {118}},\ \bibinfo {pages} {157202} (\bibinfo {year}
  {2017})}\BibitemShut {NoStop}%
\bibitem [{Note2()}]{Note2}%
  \BibitemOpen
  \bibinfo {note} {The reader will note that the right-hand side of Eq. (5)
  could be modified by a prefactor of order 1.This is why we are using an
  approximate sign in the equation, rather than an equal sign. However, the
  comparison with the simulations turns out to be satisfactory by assuming that
  the prefactor is exactly one. It is well possible that carrying out our
  program from future experiments of increased accuracy will require a more
  precise determination of this prefactor.}\BibitemShut {Stop}%
\bibitem [{Note3()}]{Note3}%
  \BibitemOpen
  \bibinfo {note} {Amusingly, although the droplets model~\cite
  {mcmillan:83,bray:87,fisher:86} and the Replica Symmetry Breaking (RSB)
  theory~\cite {marinari:00} differ in their expectation for $\theta (\xi \to
  \infty )$ (the droplets prediction is $\theta (\xi \to \infty )=0$ and
  $d=d_{\protect \text {f}}$, while RSB expects $\theta (\xi \to \infty )>0$
  and $d_{\protect \text {f}}<d$), the two theories quantitatively agree in
  their predicted behavior for $\theta (\xi )$ in our range of $\xi $~\cite
  {janus:18}.}\BibitemShut {Stop}%
\bibitem [{\citenamefont {Bouchaud}\ \emph {et~al.}(2001)\citenamefont
  {Bouchaud}, \citenamefont {Dupuis}, \citenamefont {Hammann},\ and\
  \citenamefont {Vincent}}]{bouchaud:01}%
  \BibitemOpen
  \bibfield  {author} {\bibinfo {author} {\bibfnamefont {J.-P.}\ \bibnamefont
  {Bouchaud}}, \bibinfo {author} {\bibfnamefont {V.}~\bibnamefont {Dupuis}},
  \bibinfo {author} {\bibfnamefont {J.}~\bibnamefont {Hammann}}, \ and\
  \bibinfo {author} {\bibfnamefont {E.}~\bibnamefont {Vincent}},\ }\href
  {\doibase 10.1103/PhysRevB.65.024439} {\bibfield  {journal} {\bibinfo
  {journal} {Phys. Rev. B}\ }\textbf {\bibinfo {volume} {65}},\ \bibinfo
  {pages} {024439} (\bibinfo {year} {2001})}\BibitemShut {NoStop}%
\bibitem [{\citenamefont {Berthier}\ and\ \citenamefont
  {Bouchaud}(2002)}]{berthier:02}%
  \BibitemOpen
  \bibfield  {author} {\bibinfo {author} {\bibfnamefont {L.}~\bibnamefont
  {Berthier}}\ and\ \bibinfo {author} {\bibfnamefont {J.-P.}\ \bibnamefont
  {Bouchaud}},\ }\href {\doibase 10.1103/PhysRevB.66.054404} {\bibfield
  {journal} {\bibinfo  {journal} {Phys. Rev. B}\ }\textbf {\bibinfo {volume}
  {66}},\ \bibinfo {pages} {054404} (\bibinfo {year} {2002})}\BibitemShut
  {NoStop}%
\bibitem [{Note4()}]{Note4}%
  \BibitemOpen
  \bibinfo {note} {Note that Eq.~\protect \textup {\hbox {\mathsurround \z@
  \protect \normalfont (\ignorespaces \ref {eq:well-organized-data}\unskip
  \@@italiccorr )}} correctly predicts $\protect \ensuremath {z_\protect
  \mathrm {c}}\protect \xspace (\protect \ensuremath {T_\protect \mathrm
  {c}}\protect \xspace )=6.69$, because $x=\infty $ at $\protect \ensuremath
  {T_\protect \mathrm {c}}\protect \xspace $, for all $\xi $.}\BibitemShut
  {Stop}%
\bibitem [{\citenamefont {Belletti}\ \emph {et~al.}(2008)\citenamefont
  {Belletti}, \citenamefont {Cotallo}, \citenamefont {Cruz}, \citenamefont
  {Fernandez}, \citenamefont {Gordillo-Guerrero}, \citenamefont {Guidetti},
  \citenamefont {Maiorano}, \citenamefont {Mantovani}, \citenamefont
  {Marinari}, \citenamefont {Mart\'{i}n-Mayor}, \citenamefont {Sudupe},
  \citenamefont {Navarro}, \citenamefont {Parisi}, \citenamefont
  {Perez-Gaviro}, \citenamefont {Ruiz-Lorenzo}, \citenamefont {Schifano},
  \citenamefont {Sciretti}, \citenamefont {Tarancon}, \citenamefont
  {Tripiccione}, \citenamefont {Velasco},\ and\ \citenamefont
  {Yllanes}}]{janus:08b}%
  \BibitemOpen
  \bibfield  {author} {\bibinfo {author} {\bibfnamefont {F.}~\bibnamefont
  {Belletti}}, \bibinfo {author} {\bibfnamefont {M.}~\bibnamefont {Cotallo}},
  \bibinfo {author} {\bibfnamefont {A.}~\bibnamefont {Cruz}}, \bibinfo {author}
  {\bibfnamefont {L.~A.}\ \bibnamefont {Fernandez}}, \bibinfo {author}
  {\bibfnamefont {A.}~\bibnamefont {Gordillo-Guerrero}}, \bibinfo {author}
  {\bibfnamefont {M.}~\bibnamefont {Guidetti}}, \bibinfo {author}
  {\bibfnamefont {A.}~\bibnamefont {Maiorano}}, \bibinfo {author}
  {\bibfnamefont {F.}~\bibnamefont {Mantovani}}, \bibinfo {author}
  {\bibfnamefont {E.}~\bibnamefont {Marinari}}, \bibinfo {author}
  {\bibfnamefont {V.}~\bibnamefont {Mart\'{i}n-Mayor}}, \bibinfo {author}
  {\bibfnamefont {A.~M.}\ \bibnamefont {Sudupe}}, \bibinfo {author}
  {\bibfnamefont {D.}~\bibnamefont {Navarro}}, \bibinfo {author} {\bibfnamefont
  {G.}~\bibnamefont {Parisi}}, \bibinfo {author} {\bibfnamefont
  {S.}~\bibnamefont {Perez-Gaviro}}, \bibinfo {author} {\bibfnamefont {J.~J.}\
  \bibnamefont {Ruiz-Lorenzo}}, \bibinfo {author} {\bibfnamefont {S.~F.}\
  \bibnamefont {Schifano}}, \bibinfo {author} {\bibfnamefont {D.}~\bibnamefont
  {Sciretti}}, \bibinfo {author} {\bibfnamefont {A.}~\bibnamefont {Tarancon}},
  \bibinfo {author} {\bibfnamefont {R.}~\bibnamefont {Tripiccione}}, \bibinfo
  {author} {\bibfnamefont {J.~L.}\ \bibnamefont {Velasco}}, \ and\ \bibinfo
  {author} {\bibfnamefont {D.}~\bibnamefont {Yllanes}} (\bibinfo
  {collaboration} {Janus Collaboration}),\ }\href {\doibase
  10.1103/PhysRevLett.101.157201} {\bibfield  {journal} {\bibinfo  {journal}
  {Phys. Rev. Lett.}\ }\textbf {\bibinfo {volume} {101}},\ \bibinfo {pages}
  {157201} (\bibinfo {year} {2008})} \BibitemShut
  {NoStop}%
\bibitem [{\citenamefont {Zinn-Justin}(2005)}]{zinn-justin:05}%
  \BibitemOpen
  \bibfield  {author} {\bibinfo {author} {\bibfnamefont {J.}~\bibnamefont
  {Zinn-Justin}},\ }\href@noop {} {\emph {\bibinfo {title} {Quantum Field
  Theory and Critical Phenomena}}},\ \bibinfo {edition} {4th}\ ed.\ (\bibinfo
  {publisher} {Clarendon Press},\ \bibinfo {address} {Oxford},\ \bibinfo {year}
  {2005})\BibitemShut {NoStop}%
\bibitem [{\citenamefont {McMillan}(1983)}]{mcmillan:83}%
  \BibitemOpen
  \bibfield  {author} {\bibinfo {author} {\bibfnamefont {W.~L.}\ \bibnamefont
  {McMillan}},\ }\href {\doibase 10.1103/PhysRevB.28.5216} {\bibfield
  {journal} {\bibinfo  {journal} {Phys. Rev. B}\ }\textbf {\bibinfo {volume}
  {28}},\ \bibinfo {pages} {5216} (\bibinfo {year} {1983})}\BibitemShut
  {NoStop}%
\bibitem [{\citenamefont {Bray}\ and\ \citenamefont {Moore}(1987)}]{bray:87}%
  \BibitemOpen
  \bibfield  {author} {\bibinfo {author} {\bibfnamefont {A.~J.}\ \bibnamefont
  {Bray}}\ and\ \bibinfo {author} {\bibfnamefont {M.~A.}\ \bibnamefont
  {Moore}},\ }in\ \href@noop {} {\emph {\bibinfo {booktitle} {Heidelberg
  Colloquium on Glassy Dynamics}}},\ \bibinfo {series and number} {\bibinfo
  {series} {Lecture Notes in Physics}\ No.\ \bibinfo {number} {275}},\ \bibinfo
  {editor} {edited by\ \bibinfo {editor} {\bibfnamefont {J.~L.}\ \bibnamefont
  {van Hemmen}}\ and\ \bibinfo {editor} {\bibfnamefont {I.}~\bibnamefont
  {Morgenstern}}}\ (\bibinfo  {publisher} {Springer},\ \bibinfo {address}
  {Berlin},\ \bibinfo {year} {1987})\BibitemShut {NoStop}%
\bibitem [{\citenamefont {Fisher}\ and\ \citenamefont
  {Huse}(1986)}]{fisher:86}%
  \BibitemOpen
  \bibfield  {author} {\bibinfo {author} {\bibfnamefont {D.~S.}\ \bibnamefont
  {Fisher}}\ and\ \bibinfo {author} {\bibfnamefont {D.~A.}\ \bibnamefont
  {Huse}},\ }\href {\doibase 10.1103/PhysRevLett.56.1601} {\bibfield  {journal}
  {\bibinfo  {journal} {Phys. Rev. Lett.}\ }\textbf {\bibinfo {volume} {56}},\
  \bibinfo {pages} {1601} (\bibinfo {year} {1986})}\BibitemShut {NoStop}%
\bibitem [{\citenamefont {Marinari}\ \emph {et~al.}(2000)\citenamefont
  {Marinari}, \citenamefont {Parisi}, \citenamefont {Ricci-Tersenghi},
  \citenamefont {Ruiz-Lorenzo},\ and\ \citenamefont {Zuliani}}]{marinari:00}%
  \BibitemOpen
  \bibfield  {author} {\bibinfo {author} {\bibfnamefont {E.}~\bibnamefont
  {Marinari}}, \bibinfo {author} {\bibfnamefont {G.}~\bibnamefont {Parisi}},
  \bibinfo {author} {\bibfnamefont {F.}~\bibnamefont {Ricci-Tersenghi}},
  \bibinfo {author} {\bibfnamefont {J.~J.}\ \bibnamefont {Ruiz-Lorenzo}}, \
  and\ \bibinfo {author} {\bibfnamefont {F.}~\bibnamefont {Zuliani}},\ }\href
  {\doibase 10.1023/A:1018607809852} {\bibfield  {journal} {\bibinfo  {journal}
  {J. Stat. Phys.}\ }\textbf {\bibinfo {volume} {98}},\ \bibinfo {pages} {973}
  (\bibinfo {year} {2000})}
  \BibitemShut {NoStop}%
\end{thebibliography}

\end{document}